\documentclass[twocolumn,showpacs,preprintnumbers,amsmath,amssymb]{revtex4}

\usepackage{graphicx}
\usepackage{dcolumn}
\usepackage{bm}

\begin{document}

\preprint{}

\title{Anomalous Hall Effect due to Non-collinearity \\
in Pyrochlore Compounds: Role of Orbital Aharonov-Bohm Effect}

\author{Takeshi Tomizawa}

\author{Hirhoshi Kontani}
\affiliation{%
Department of Physics, Nagoya University, Furo-cho, Nagoya 464-8602, Japan.
}%

\date{\today}

\begin{abstract}
To elucidate the origin of spin structure-driven anomalous Hall effect (AHE) 
in pyrochlore compounds, we construct the $t_{2g}$-orbital kagome lattice model
and analyze the anomalous Hall conductivity (AHC).
We reveal that a conduction electron acquires a Berry phase 
due to the complex $d$-orbital wavefunction
in the presence of spin-orbit interaction.
This ``orbital Aharonov-Bohm (AB) effect'' produces the AHC 
that is drastically changed in the presence of 
non-collinear spin structure.
In both ferromagnetic compound $\rm Nd_2Mo_2O_7$
and paramagnetic compound $\rm Pr_2Ir_2O_7$,
the AHC given by the orbital AB effect
totally dominates the spin chirality mechanism,
and succeeds in explaining the experimental relation between the 
spin structure and the AHC.
Especially, ``finite AHC in the absence of magnetization''
observed in $\rm Pr_2Ir_2O_7$ can be explained 
in terms of the orbital mechanism
by assuming small magnetic order of Ir $5d$-electrons.

\end{abstract}

\maketitle

\section{\label{sec:level1}Introduction}

\begin{figure}
\includegraphics[scale=0.2]{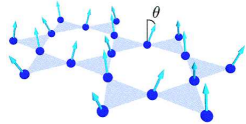}
\caption{\label{fig:TiltedFerro} Tilted ferromagnetic state in the kagome lattice. 
Blue circles are Mo ions. Arrows at Mo sites are the tilted ferromagnetic exchange field. }
\end{figure}

Recently, theory of intrinsic anomalous Hall effect (AHE)
in multiband ferromagnetic metals has been developed 
intensively from the original work by Karplus and Luttinger (KL)
\cite{KarplusLuttinger}.
The anomalous Hall conductivity (AHC) 
$\sigma_{\rm AH} \equiv j_x/E_y$ due to intrinsic AHE
shows the almost material-specific value that is independent of 
the relaxation time.
The intrinsic AHE in heavy fermion compounds \cite{KontaniYamada}, 
Fe \cite{Yao}, and Ru-oxides \cite{Miyazawa,Fang,KontaniTanakaYamada}
had been studied intensively based on realistic multiband models. 
Also, large spin Hall effect (SHE) 
observed in Pt and other paramagnetic transition metals \cite{Kimura},
which is analog to the AHE in ferromagnets,
is also reproduced well in terms of the intrinsic Hall effect
 \cite{KontaniTanakaHirashima,Guo,TanakaKontani}.
The intrinsic AHE and SHE in transition metals
originate from the Berry phase given by the $d$-orbital angular momentum
induced by the spin-orbit interaction (SOI), 
which we call the ``orbital Aharonov-Bohm (AB) effect'' \cite{KontaniTanaka2}.

In particular, AHE due to nontrivial spin structure
attracts increasing attention,
such as Mn oxides \cite{Ye} and spin glass systems \cite{Tatara}.
The most famous example would be the pyrochlore compound $\rm Nd_2Mo_2O_7$
 \cite{Yoshii,Kageyama,Yasui,Taguchi}.
Here, Mo 4$d$ electrons are in the ferromagnetic state below $T_{\rm c}=93$K,
and the tilted ferromagnetic state in Fig. \ref{fig:TiltedFerro}
is realized by the non-coplanar Nd 4$f$ magnetic order below 
$T_{\rm N}\approx30$K, due to the $d$-$f$ exchange interaction. 
Below $T_{\rm N}$, the AHC is drastically changed by 
the small change in the tilting angle $\theta$ of Mo spin;
$\theta^2<10^{-3}$ in the neutron-diffraction study
 \cite{Yasui}.
This behavior strongly deviates from the KL-type
conventional behavior $\sigma_{AH}\propto M_z \propto 1-\frac12 \theta^2$.
Moreover, the AHC given by the spin chirality mechanism 
\cite{Ohgushi,Taillefumier}, 
which is proportional to the solid angle 
$\bm{s}_{\rm A}\cdot(\bm{s}_{\rm B}\times\bm{s}_{\rm C})\propto\theta^2$ 
subtended by three spins,
is also too small to explain experiments.
Moreover, $\langle\theta^2\rangle$ takes the minimum value
under $H\sim3$ Tesla according to the neutron-diffraction study \cite{Yasui},
whereas the AHC monotonically decreases with $H$.
Thus, the origin of the unconventional AHE in 
$\rm Nd_2Mo_2O_7$ had been an open problem for a long time.

Very recently, this problem was revisited by the present authors
by considering the $d$-orbital degree of freedom and the atomic SOI
 \cite{TomizawaKontani},
and found that a drastic spin structure-driven AHE emerges
due to the orbital AB effect, in the presence of non-collinear spin order.
Since the obtained AHC is linear in $\theta$, 
it is much larger than the spin chirality term for $|\theta|\ll1$. 
In Ref. \cite{TomizawaKontani}, we constructed the $t_{2g}$ 
orbital kagome lattice model based on the spinel structure ($X{\rm Mo_2 O_4}$):
Although Mo atoms in $X{\rm Mo_2 O_4}$ and $X_2{\rm Mo_2 O_7}$ 
are equivalent in position and forms the pyrochlore lattice, 
positions of O atoms in $X_2{\rm Mo_2 O_7}$ are much complicated.

In this paper, we construct the $t_{2g}$ kagome lattice tight-biding model 
based on the pyrochlore structure, 
by taking the crystalline electric field into account. 
We find that the orbital AB effect causes large $\theta$-linear AHC,
resulting from the combination of the non-collinear spin order 
(including orders with zero scalar chirality) and atomic SOI. 
The realized AHC is much larger than the spin chirality term
due to non-coplanar spin order, and it explains the salient features 
of spin structure-driven AHE in $\rm Nd_2Mo_2O_7$. 
We also study another pyrochlore compound $\rm Pr_2Ir_2O_7$,
and find that the orbital AB effect also gives the dominant contribution:
We show that important features of the unconventional AHE in 
$\rm Pr_2Ir_2O_7$, such as highly non-monotonic field dependence and 
residual AHC in the absence of magnetization, 
are well reproduced by the orbital AB effect.

The paper is organized as follows:
In Sec. II, we introduce the pyrochlore-type $t_{2g}$ orbital 
tight-binding model and the Hamiltonian. 
We give the general expressions for the intrinsic AHC 
in Sec. III, and explain the orbital Aharonov-Bohm effect in Sec. IV.
The numerical results for $\rm Nd_2Mo_2O_7$ and $\rm Pr_2Ir_2O_7$
are presented in Sec. V and VI, respectively.
In Sec. VII, we make comparison between theory and experiment.

\section{Model and Hamiltonian}

\begin{figure}
\includegraphics[scale=0.23]{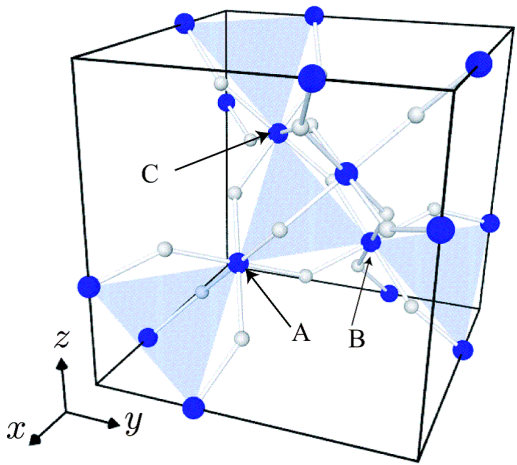}
\caption{\label{fig:pyro} Pyrochlore structure. 
Blue (white) circles are Mo (O) ions. 
The Mo ions on the [111] plane form the kagome lattice.}
\end{figure}

\begin{figure}
\includegraphics[scale=0.23]{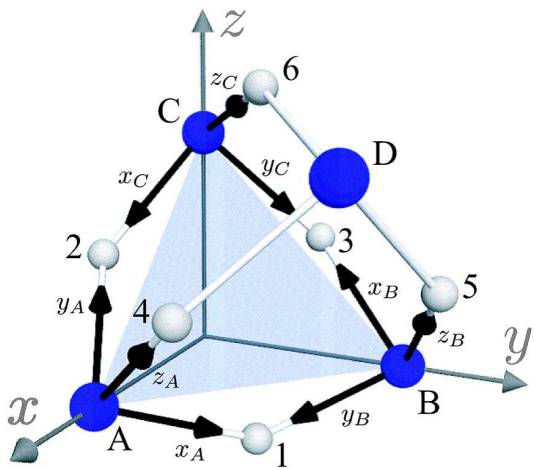}
\caption{\label{fig:pyroxyz2} Configurations of Mo$_{\rm A-D}$ and O$_{1-6}$. 
The $x_\Xi y_\Xi z_\Xi$-coordinate is defined by the surrounding O
tetrahedron. }
\end{figure}

\begin{table}
\caption{\label{tab:table1} Coordinates of Mo and O in pyrochlore structure as shown in Fig. \ref{fig:pyroxyz2} in the $xyz$-coordinate.}
\label{tab:coordinates}
\begin{ruledtabular}
\begin{tabular}{lll}
Ion&Site&Coordinate\\
\hline
Mo&A& (1/4 ,0, 0) \\
  &B& (0, 1/4, 0) \\
  &C& (0, 0, 1/4) \\
  &D& (1/4, 1/4, 1/4) \\
\hline
O&1& (1/8, 1/8, -1/16) \\
 &2& (1/8, -1/16, 1/8) \\
 &3& (-1/16, 1/8, 1/8) \\
 &4& (5/16, 1/8, 1/8) \\
 &5& (1/8, 5/16, 1/8) \\
 &6& (1/8, 1/8, 5/16) \\
\end{tabular}
\end{ruledtabular}
\end{table}
 
First, we introduce the crystal structure of the pyrochlore oxide 
$\rm A_2B_2O_7$: It has the face centered cubic structure, 
in which two individual 3-dimensional networks of the corner-sharing 
A$_4$ and B$_4$ tetrahedron are formed. 
In this paper, we mainly discuss the AHE in $\rm Nd_2Mo_2O_7$, 
and $\rm Pr_2Ir_2O_7$ is also discussed in section \ref{section: Pr2Ir2O7}. 
Figure \ref{fig:pyro} represents the Mo ions (Blue circles) and O ions (White circles) in the pyrochlore structure.
The [111] Mo layer forms the kagome lattice.
The Mo $4d$-electrons give itinerant carriers while the Nd $4f$-electrons
form local moments.

We construct pyrochlore type $t_{2g}$-orbital tight binding model 
in the kagome lattice for Mo $4d$ electrons, where the unit cell 
contains three sites A, B and C in Fig. \ref{fig:pyroxyz2}. 
The coordinates of Mo and O are shown in Table \ref{tab:coordinates}
\cite{Subramanian}, and the quantization axis for the Mo $d$-orbital 
is fixed by the surrounding O$_6$ octahedron. 
To describe the $d$-orbital state, we introduce the $(xyz)_\Xi$-coordinate for $\Xi={\rm A,B,C}$ sites shown by Fig.\ref{fig:pyroxyz2}. 
The $(xyz)_\Xi$-coordinate is defined by the surrounding O$_6$ ions. 
In the case of the $(xyz)_{\rm A}$-coordinate, we choose $x_{\rm A}$, $y_{\rm A}$ and  $z_{\rm A}$ axes 
as Mo$_{\rm A}\rightarrow$O$_1$, Mo$_{\rm A}\rightarrow$O$_2$ and 
Mo$_{\rm A}\rightarrow$O$_4$ direction, respectively, in Fig. \ref{fig:pyroxyz2}. 
We also choose the $(xyz)_{\rm B}$- and $(xyz)_{\rm C}$-coordinates in the same way. 
\begin{figure}
\includegraphics[scale=0.19]{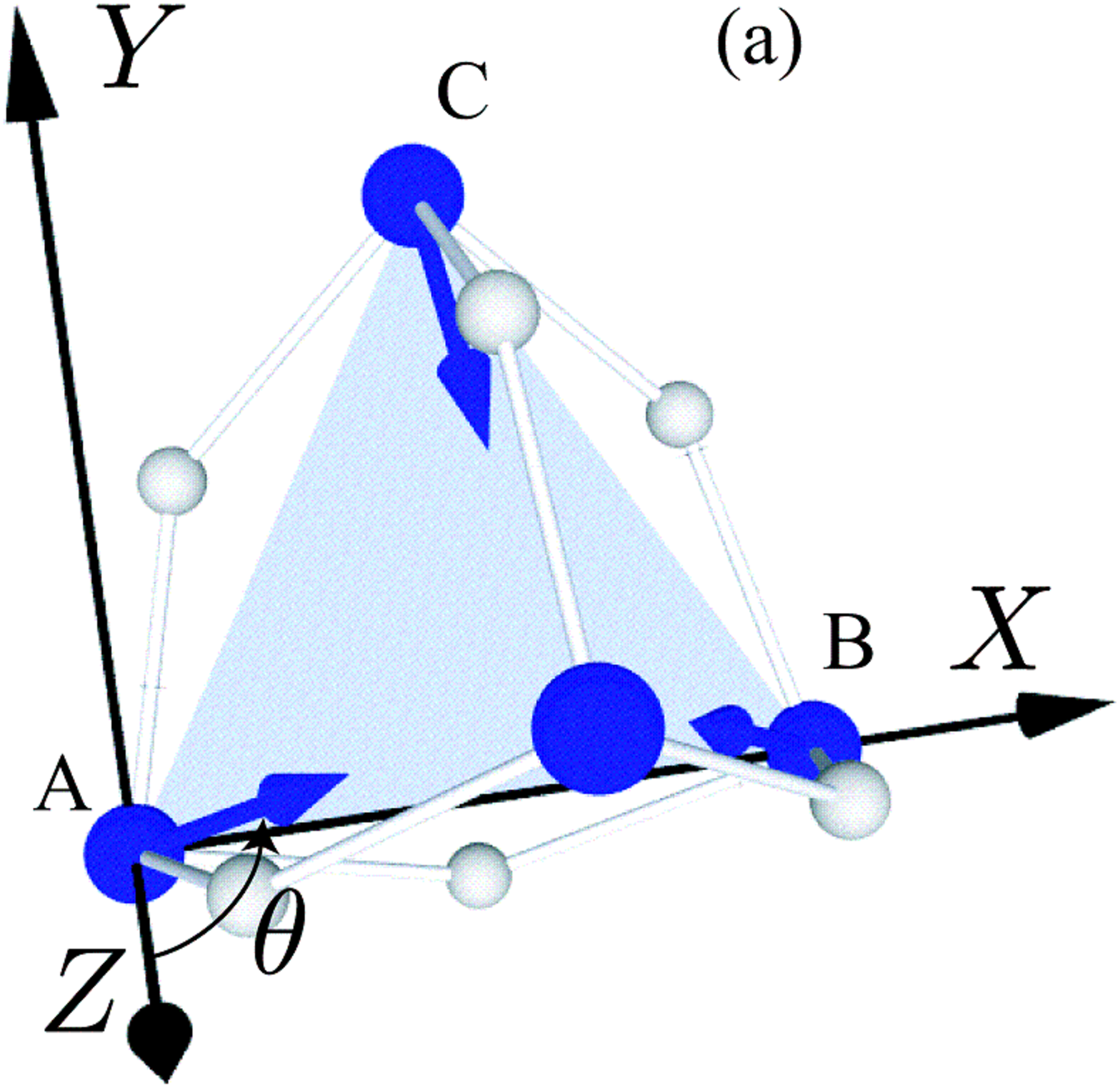}

\includegraphics[scale=0.3]{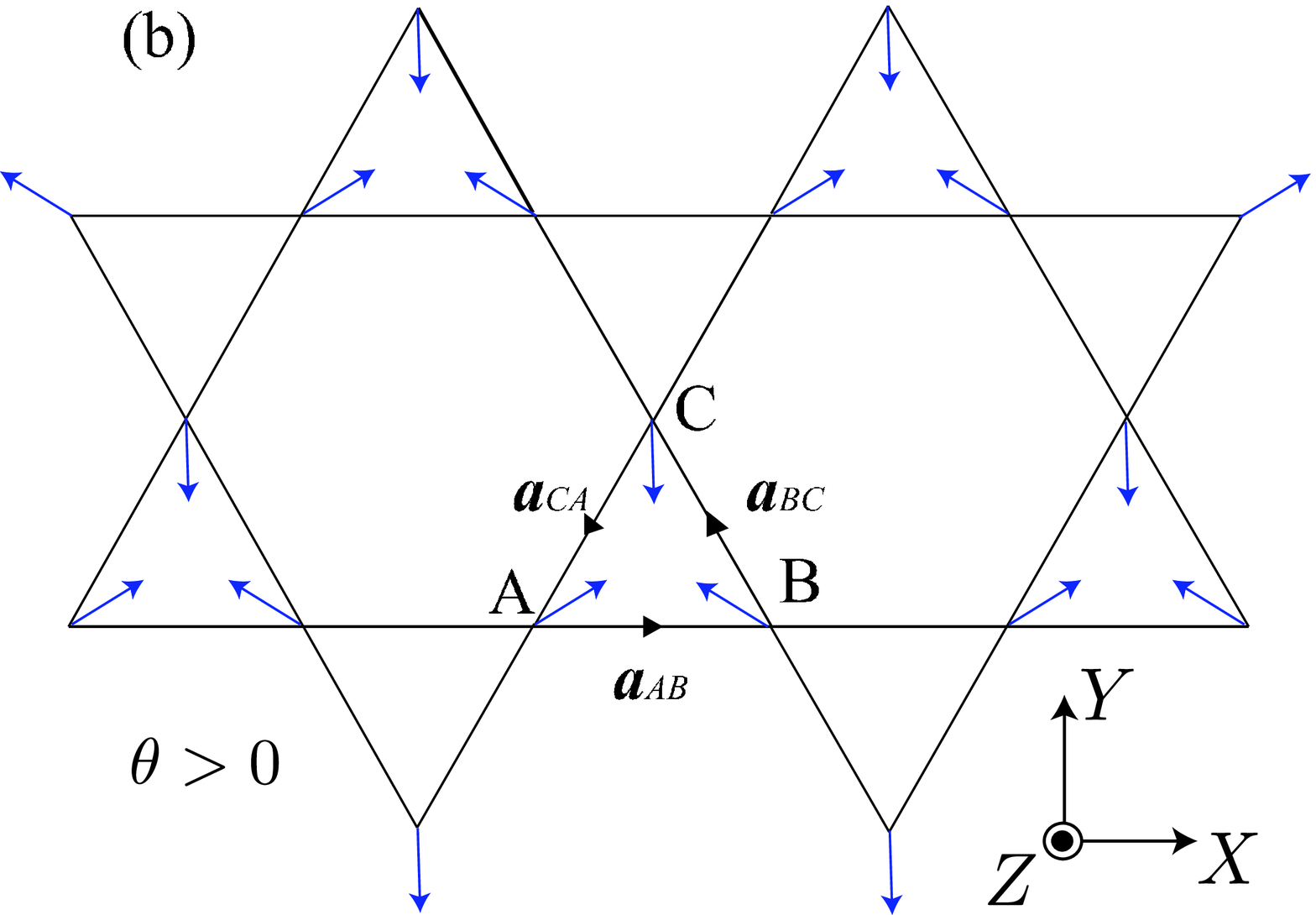}
\includegraphics[scale=0.3]{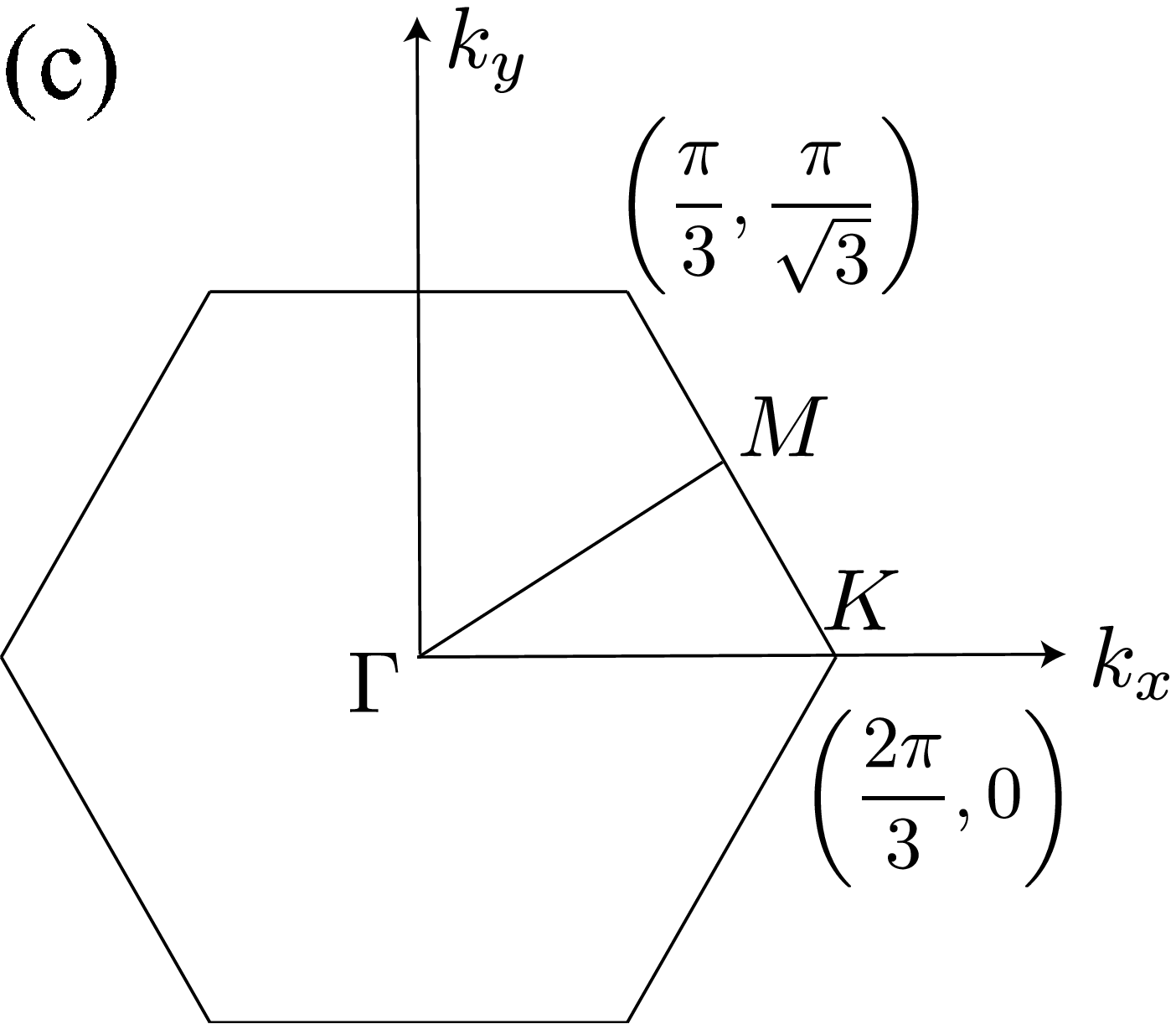}
\caption{\label{fig:pyroXYZ} 
(a) Umbrella like locale exchange field at Mo sites represented by arrows. 
A unit cell contains sites A, B, C. 
(b) Kagome lattice. ${\boldsymbol a}_{ij}\ (i,j=A,B,C)$ is 
a half Bravais vector. 
(c) First Brillouin zone in a hexagon shape.}
\end{figure}

Moreover, we introduce the $XYZ$-coordinate on the kagome layer shown in Fig. \ref{fig:pyroXYZ}(b).
We choose $X$ axis as Mo$_A\rightarrow$Mo$_B$ direction and 
$Y$ axis is perpendicular to $X$ axis on the kagome layer. 
A vector $(n_{x},n_{y},n_{z})_\Xi$ in the  $(xyz)_\Xi$-coordinate
is transformed into $[n_X,n_Y,n_Z]$ in the $XYZ$-coordinate as 
$(n_{x},n_{y},n_{z})_\Xi=[n_X,n_Y,n_Z]\hat{O}_\Xi$, where the coordinate transformation matrix $\hat{O}_\Xi$ is given by

\begin{subequations}
\label{eq:O}
\begin{eqnarray}
\hat{O}_{\rm A}&=&\frac{1}{3\sqrt{6}}
\begin{pmatrix}
4\sqrt{3}&\sqrt{3}&\sqrt{3}\\
-2&7&1\\
-\sqrt{2}&-\sqrt{2}&5\sqrt{2}
\end{pmatrix},\label{eq:OA}\\
\hat{O}_{\rm B}&=&\frac{1}{3\sqrt{6}}
\begin{pmatrix}
-\sqrt{3}&-4\sqrt{3}&-\sqrt{3}\\
7&-2&1\\
-\sqrt{2}&-\sqrt{2}&5\sqrt{2}
\end{pmatrix},\label{eq:OB}\\
\hat{O}_{\rm C}&=&\frac{1}{3\sqrt{6}}
\begin{pmatrix}
-3\sqrt{3}&3\sqrt{3}&0\\
-5&-5&-2\\
-\sqrt{2}&-\sqrt{2}&5\sqrt{2}
\end{pmatrix}\label{eq:OC}.
\end{eqnarray}
\end{subequations}

Arrows in Fig. \ref{fig:pyroXYZ}(a) represents the local effective magnetic field at Mo sites, 
which is composed of the ferromagnetic exchange field for Mo 4$d$-electrons and the exchange field from Nd 4$f$ electrons.
Under the magnetic field parallel to $[111]$ direction below $T_N$, 
the direction of the local exchange fields at sites A, B and C in the
$XYZ$-coordinate are 
$(\phi_{\rm A} =\pi/6,\theta)$, $(\phi_{\rm B} =5\pi/6,\theta)$ and 
$(\phi_{\rm C} =-\pi/2,\theta)$, respectively. 
In Nd$_2$Mo$_2$O$_7$,
the tilting angle $\theta$ changes from negative to positive
as $H$ increases from $+0$ Tesla, corresponding to the change in the 
spin-ice state at Nd sites \cite{Yasui,Sato}.

Now, we explain the Hamiltonian. 
The Hamiltonian for the $t_{2g}$-orbital kagome lattice
model is given by
\begin{eqnarray}
H=&&\sum_{i\alpha,j\beta,\sigma}t_{i\alpha,j\beta}c^\dagger_{i\alpha,\sigma}c_{j\beta,\sigma}
-\sum_{i\alpha,\sigma\sigma'}{\boldsymbol h}_i\cdot[{\boldsymbol \mu}_e]_{\sigma,\sigma'}c^\dagger_{i\alpha,\sigma}c_{i\alpha,\sigma'}\nonumber\\
&&+\lambda\sum_{i\alpha\beta,\sigma\sigma'}[{\boldsymbol l}]_{\alpha,\beta}\cdot[\mathbf{s}]_{\sigma,\sigma'}c^\dagger_{i\alpha,\sigma}c_{j\beta,\sigma}
\label{eq:Hamiltonian}
\end{eqnarray}
where $c^\dagger$ is a creation operator for $4d$-electron on Mo ions
while the field ${\boldsymbol h}$ arises from the
ordered Nd moments, which are treated as a static, classical background.
$(i, j)$, $(\alpha, \beta)$ and $(\sigma, \sigma')$ represent the sites, $t_{2g}$-orbitals and spins, respectively. 
Hereafter, we denote the $t_{2g}$-orbitals $(xy,yz,zx)$ as $(1,2,3)$ for simplicity. 
The first term in eq. (\ref{eq:Hamiltonian})
describes electrons hopping.
$t_{i\alpha,j\beta}$ is the hopping integrals between $(i, \alpha)$ and $(j, \beta)$. 
The direct $d$-$d$ hopping integrals are given by the Slater-Koster (SK) parameters 
$(dd\sigma)$, $(dd\pi)$ and $(dd\delta)$ \cite{SlaterKoster}.
In the present model, however, SK parameter table given in Ref. \cite{SlaterKoster} is not available 
since the $d$-orbitals at each site are described in the different coordinate as shown in Fig. \ref{fig:pyroxyz2}. 
In Appendix A, we will derive the hopping integral between the sites with different coordinates. 
The second term in eq. (\ref{eq:Hamiltonian})
represents the Zeeman term, where ${\boldsymbol h}_i$ is the local exchange field at site $i$.
${\boldsymbol \mu}_e\equiv -2{\boldsymbol s}$ is the magnetic moment of an electron. 
Here, we put $\mu_B$=1. 
The third term represents the SOI, where $\lambda$ is the spin-orbit coupling constant, 
and ${\boldsymbol l}$ and ${\boldsymbol s}$ are the $d$-orbital and spin operators, respectively.

The Hamiltonian in Eq. (\ref{eq:Hamiltonian}) is rewritten 
in the momentum space as 
\begin{equation}
H=\sum_{\boldsymbol k}C^\dagger_{\boldsymbol k}\hat{H}_{\boldsymbol k}C_{\boldsymbol k},
\label{eq:Hamiltonian2}
\end{equation}
where $\boldsymbol k$ summation is over the first Brillouin zone in Fig. \ref{fig:pyroXYZ}(c), and
\begin{eqnarray}
C_{\boldsymbol k}^\dagger=
(
{a^\dagger_{{\boldsymbol k},1 \uparrow}},{a^\dagger_{{\boldsymbol k},1 \downarrow}},{b^\dagger_{{\boldsymbol k},1 \uparrow}},
{b^\dagger_{{\boldsymbol k},1 \downarrow}},{c^\dagger_{{\boldsymbol k},1 \downarrow}},{c^\dagger_{{\boldsymbol k},1 \uparrow}}\nonumber\\
{a^\dagger_{{\boldsymbol k},2 \uparrow}},{a^\dagger_{{\boldsymbol k},2 \downarrow}},{b^\dagger_{{\boldsymbol k},2 \uparrow}},
{b^\dagger_{{\boldsymbol k},2 \downarrow}},{c^\dagger_{{\boldsymbol k},2 \downarrow}},{c^\dagger_{{\boldsymbol k},2 \uparrow}}\nonumber\\
{a^\dagger_{{\boldsymbol k},3 \uparrow}},{a^\dagger_{{\boldsymbol k},3 \downarrow}},{b^\dagger_{{\boldsymbol k},3 \uparrow}},
{b^\dagger_{{\boldsymbol k},3 \downarrow}},{c^\dagger_{{\boldsymbol k},3 \downarrow}},{c^\dagger_{{\boldsymbol k},3 \uparrow}}
). 
\end{eqnarray}
Here and hereafter, we denote the creation operators at sites A, B and C 
as $a^\dagger_{{\boldsymbol k},\alpha \sigma}, b^\dagger_{{\boldsymbol k},\alpha \sigma}$ and 
$c^\dagger_{{\boldsymbol k},\alpha \sigma}$, respectively. 
$\hat{H}_{\boldsymbol k}$ is given by 18$\times$18 matrix: 

\begin{equation}
\hat{H}_{\boldsymbol k}
=
\begin{pmatrix}
\hat{H}_{{\boldsymbol k}11}&\hat{H}_{{\boldsymbol k}12}&\hat{H}_{{\boldsymbol k}13}\\
\hat{H}_{{\boldsymbol k}21}&\hat{H}_{{\boldsymbol k}22}&\hat{H}_{{\boldsymbol k}23}\\
\hat{H}_{{\boldsymbol k}31}&\hat{H}_{{\boldsymbol k}32}&\hat{H}_{{\boldsymbol k}33}
\end{pmatrix},
\end{equation}
where $\hat{H}_{{\boldsymbol k}\alpha\beta}$ is a 6$\times$6 matrix
with respect to $(\Xi,\sigma)$.

Here, we divide the Hamiltonian (\ref{eq:Hamiltonian2}) into four parts: 
\begin{equation}
\hat{H}_{{\boldsymbol k}\alpha\beta}=\hat{H}^{t}_{{\boldsymbol k}\alpha\beta}
+\hat{H}^{\rm Ze}_{\alpha\beta}
+\hat{H}^{\lambda}_{\alpha\beta}
+\hat{H}^{\rm CEF}_{\alpha\beta},\label{eq:Hktot}
\end{equation}
where we added the crystalline electric field potential term $\hat{H}^{\rm CEF}_{\alpha\beta}$ 
to Eq. (\ref{eq:Hamiltonian}).
The kinetic term $\hat{H}^{t}_{{\boldsymbol k}\alpha\beta}$ is given by
\begin{widetext}
\begin{eqnarray}
\hat{H}_{{\boldsymbol k}\alpha\beta}^{t}=
\bordermatrix{
                   &\beta {\rm A}\uparrow   &\beta {\rm A}\downarrow &\beta {\rm B}\uparrow   &\beta {\rm B}\downarrow &\beta {\rm C}\uparrow   &\beta {\rm C}\downarrow\cr
 \alpha {\rm A}\uparrow  &0                 &0                 &p_{{\rm A}\alpha,{\rm B}\beta}&0                 &p_{{\rm A}\alpha,{\rm C}\beta}&0\cr
 \alpha {\rm A}\downarrow&0                 &0                 &0                 &p_{{\rm A}\alpha,{\rm B}\beta}&0                 &p_{{\rm A}\alpha,{\rm C}\beta}\cr
 \alpha {\rm B}\uparrow  &p_{{\rm B}\alpha,{\rm A}\beta}&0                 &0                 &0                 &p_{{\rm B}\alpha,{\rm C}\beta}&\cr
 \alpha {\rm B}\downarrow&0                 &p_{{\rm B}\alpha,{\rm A}\beta}&0                 &0                 &0                 &p_{{\rm B}\alpha,{\rm C}\beta}\cr
 \alpha {\rm C}\uparrow  &p_{{\rm C}\alpha,{\rm A}\beta}&0                 &p_{{\rm C}\alpha,{\rm B}\beta}&0                 &0                 &0\cr
 \alpha {\rm C}\downarrow&0                 &p_{{\rm C}\alpha,{\rm A}\beta}&0                 &p_{{\rm C}\alpha,{\rm B}\beta}&0                 &0  
},
\end{eqnarray}
\end{widetext}
where $p_{i\alpha,j\beta}=2t_{i\alpha,j\beta}\cos({\boldsymbol k}\cdot{\boldsymbol a_{ij}})$ and ${\boldsymbol a}_{ij}$ is a half Bravais vector 
in Fig. \ref{fig:pyroXYZ}(b). 

The Zeeman term $\hat{H}^{\rm Ze}_{\alpha\beta}$ is given by \cite{Taillefumier}
%
%
\begin{eqnarray}
\hat{H}_{\alpha\beta}^{\rm Ze}&=&h_0\delta_{\alpha\beta}
\bordermatrix{
         &\beta{\rm A}&\beta{\rm B}&\beta{\rm C}\cr
\alpha{\rm A}  & {\hat R}_{\theta,\pi/6} &     0     &0\cr
\alpha{\rm B}  &    0     & {\hat R}_{\theta,5\pi/6}  &0\cr
\alpha{\rm C}  &    0     &   0       &{\hat R}_{\theta,-\pi/2}\cr
}, \\
\hat{R}_{\theta,\phi}&=&
\bordermatrix{
                  &\uparrow&\downarrow\cr
\uparrow  & \cos\theta & \sin\theta e^{-i\phi}    \cr
\downarrow  & \sin\theta e^{i\phi} & -\cos\theta \cr
}, 
\end{eqnarray}
where $h_0$=$|{\boldsymbol h}_i|$,
and $\delta_{\alpha\beta}$ is a Kronecker's delta.

Now, we consider the SOI term $\hat{H}^{\lambda}$ 
in Eq. (\ref{eq:Hktot}). 
For convenience in calculating the AHC, we take the $Z$-axis
for the spin quantization axis.
Then, $2\bm{s}=2[s_X,s_Y,s_Z]$ is given by the Pauli matrix vector
in the $XYZ$-coordinate.
To derive the $\hat{H}^{\lambda}_{\alpha\beta}$, however,
we have to express the spin operator in the $(xyz)_\Xi$-coordinate,
which is given by the relationship
$(s_{x}^{\Xi},s_{y}^{\Xi},s_{z}^{\Xi})=[s_X,s_Y,s_Z]\hat{O}_\Xi$
and Eqs. (\ref{eq:OA})-(\ref{eq:OC}).
In the $(xyz)_\Xi$-coordinate, the nonzero matrix elements of $\bm l$ 
are given as
$\langle 3|l_x|1\rangle=\langle 1|l_y|2\rangle=\langle 2|l_z|3\rangle=i$
and their Hermite conjugates \cite{Friedel, TanakaKontani}.
Thus, the matrix elements
$({\hat H}_{\alpha,\beta}^{\lambda})_{\Xi\sigma,\Xi'\sigma'}\equiv 
\langle\Xi\alpha\sigma|{\hat H}^{\lambda}|\Xi\beta\sigma'\rangle 
\cdot \delta_{\Xi,\Xi'}$
for $(\alpha,\beta)=(3,1)$ are given as
\begin{subequations}
 \begin{eqnarray}
({\hat H}_{3,1}^{\lambda})_{{\rm A}\sigma,{\rm A}\sigma'}
\!\!&=&\!\!\frac{i\lambda}{3\sqrt{6}}\langle \sigma|(4\sqrt{3}s_X-2s_Y-\sqrt{2}s_Z)|\sigma'\rangle,
 \nonumber \\
({\hat H}_{3,1}^{\lambda})_{{\rm B}\sigma,{\rm B}\sigma'}
\!\!&=&\!\!\frac{i\lambda}{3\sqrt{6}}\langle \sigma|(-\sqrt{3}s_X+7s_Y-\sqrt{2}s_Z)|\sigma'\rangle,
 \nonumber \\
({\hat H}_{3,1}^{\lambda})_{{\rm C}\sigma,{\rm C}\sigma'}
\!\!&=&\!\!\frac{i\lambda}{3\sqrt{6}}\langle \sigma|(-3\sqrt{3}s_X-5s_Y-\sqrt{2}s_Z)|\sigma'\rangle .
 \nonumber
 \end{eqnarray}
\end{subequations}
Thus, the $3$-$1$ component of the third term in Eq. (\ref{eq:Hktot}) becomes

\begin{widetext}
\begin{equation}
\hat{H}^\lambda_{31}=\frac{i\lambda}{6\sqrt{6}}\times
\bordermatrix{
            &1{\rm A}\uparrow  &1{\rm A}\downarrow&1{\rm B}\uparrow  &1{\rm B}\downarrow&1{\rm C}\uparrow   &1{\rm C}\downarrow\cr
3{\rm A}\uparrow  &-\sqrt{2}   &4\sqrt{3}+2i&0           &0           &0            &0\cr
3{\rm A}\downarrow&4\sqrt{3}-2i&\sqrt{2}    &0           &0           &0            &0\cr
3{\rm B}\uparrow  &0           &0           &-\sqrt{2}   &-\sqrt{3}-7i&0            &0\cr
3{\rm B}\downarrow&0           &0           &-\sqrt{3}+7i&\sqrt{2}    &0            &0\cr
3{\rm C}\uparrow  &0           &0           &0           &0           &-\sqrt{2}    &-3\sqrt{3}+5i\cr
3{\rm C}\downarrow&0           &0           &0           &0           &-3\sqrt{3}-5i&\sqrt{2}
}.
\end{equation}
The $1$-$2$ and $2$-$3$ components are calculated in a similar way. 
The obtained results are given by
\begin{eqnarray}
\hat{H}^\lambda_{12}&=&\frac{i\lambda}{6\sqrt{6}}\times
\bordermatrix{
            &2{\rm A}\uparrow  &2{\rm A}\downarrow&2{\rm B}\uparrow   &2{\rm B}\downarrow &2{\rm C}\uparrow   &2{\rm C}\downarrow\cr
1{\rm A}\uparrow  &-\sqrt{2}   &\sqrt{3}-7i &0            &0            &0            &0\cr
1{\rm A}\downarrow&\sqrt{3}+7i &\sqrt{2}    &0            &0            &0            &0\cr
1{\rm B}\uparrow  &0           &0           &-\sqrt{2}    &-4\sqrt{3}+2i&0            &0\cr
1{\rm B}\downarrow&0           &0           &-4\sqrt{3}-2i&\sqrt{2}     &0            &0\cr
1{\rm C}\uparrow  &0           &0           &0            &0            &-\sqrt{2}    &3\sqrt{3}+5i\cr
1{\rm C}\downarrow&0           &0           &0            &0            &3\sqrt{3}-5i &\sqrt{2}
},\\
\hat{H}^\lambda_{23}&=&\frac{i\lambda}{6\sqrt{6}}\times
\bordermatrix{
            &3{\rm A}\uparrow  &3{\rm A}\downarrow&3{\rm B}\uparrow  &3{\rm B}\downarrow&3{\rm C}\uparrow   &3{\rm C}\downarrow\cr
2{\rm A}\uparrow  &5\sqrt{2}  &\sqrt{3}-i  &0           &0           &0            &0\cr
2{\rm A}\downarrow&\sqrt{3}+i  &-5\sqrt{2}   &0           &0           &0            &0\cr
2{\rm B}\uparrow  &0           &0           &5\sqrt{2}  &-\sqrt{3}-i &0            &0\cr
2{\rm B}\downarrow&0           &0           &-\sqrt{3}+i &-5\sqrt{2}   &0            &0\cr
2{\rm C}\uparrow  &0           &0           &0           &0           &5\sqrt{2}   &2i\cr
2{\rm C}\downarrow&0           &0           &0           &0           &-2i          &-5\sqrt{2}
}.
\end{eqnarray}
\end{widetext}

Finally, we consider the crystalline electric field Hamiltonian $\hat{H}^{\rm CEF}_{\alpha\beta}$, 
which describes the splitting of $t_{2g}$ level into two levels $a_{1g}$ (non-degeneracy) and $e'_g$ (two-fold degeneracy) by the trigonal
deformation of MoO$_6$ octahedron.
The crystalline electric field Hamiltonian in this case is given by
\begin{eqnarray}
\hat{H}^{\rm CEF}_{\alpha\beta}=E_0(1-\delta_{\alpha,\beta})\cdot {\hat 1}
\end{eqnarray}
The eigenvalues of $\hat{H}^{\rm CEF}$ at each site are $2E_0$ for $a_{1g}$ state; 
$|a_{1g}\rangle= \frac1{\sqrt{3}}(|xy\rangle+|yz\rangle+|zx\rangle)$,  
and $-E_0$ for $e'_g$ states;
$|e_g^1\rangle= \frac1{\sqrt{2}}(|yz\rangle-|zx\rangle)$ and 
$|e_g^2\rangle= \frac1{\sqrt{6}}(2|xy\rangle-|yz\rangle-|zx\rangle)$.
Thus, the crystalline electric field splitting between $a_{1g}$ and $e'_g$ is $3|E_0|$.


\section{Anomalous Hall conductivity}
In this section, we propose the general expressions for the intrinsic AHC based on the linear-response theory.
The Green function is given by a $18\times18$ matrix: 
$\hat{G}_{\boldsymbol k}(\epsilon)=((\epsilon+\mu)\hat{1}-\hat{H}_{\boldsymbol k})^{-1}$, 
where $\mu$ is the chemical potential.
According to the linear response theory, the AHC is given by 
$\sigma_{\rm AH}=\sigma^{\rm I}_{\rm AH}+\sigma^{\rm II}_{\rm AH}$ \cite{Streda}:
\begin{eqnarray}
\sigma^{\rm I}_{\rm AH} &&=\frac{1}{2\pi N}\sum_{\boldsymbol k}{\rm Tr}\left[\hat{j}_X\hat{G}^R\hat{j}_Y\hat{G}^A\right]_{\epsilon=0}\label{eq:sigma1}\\
\sigma^{\rm II}_{\rm AH} &&=\frac{-1}{4\pi N}\sum_{\boldsymbol k}\int_{-\infty}^\mu d\epsilon 
{\rm Tr}\Biggl[\hat{j}_X\frac{\partial\hat{G}^R}{\partial \epsilon}\hat{j}_Y\hat{G}^A\nonumber\\
&&-\hat{j}_X\hat{G}^R\hat{j}_Y\frac{\partial\hat{G}^A}{\partial \epsilon}- \langle R\rightarrow A\rangle \Biggr]. \label{eq:sigma2}
\end{eqnarray}
Here, $\hat{G}_{\boldsymbol k}^{R(A)}(\epsilon)\equiv\hat{G}_{\boldsymbol k}(\epsilon+(-)i\gamma)$ 
is the retarded (advanced) Green function, 
where $\gamma (>0)$ is the quasiparticle damping rate. 
$\hat{j}_{{\boldsymbol k}\mu}\equiv -e\partial \hat{H}_{\boldsymbol k} / \partial k_\mu =-e\hat{v}_\mu \ (\mu= X,Y)$
is the charge current, where $-e$ is the electron charge.
Since all the matrix $\hat{j}_{{\boldsymbol k}\mu}$ is odd 
with respect to $\boldsymbol k$, the current vertex correction due to local impurities vanishes identically \cite{TanakaKontani, KontaniTanakaHirashima}. 
Thus, we can safely neglect the current vertex correction in calculating AHC in the present model. 
In the band-diagonal representation, eqs. (\ref{eq:sigma1}) and (\ref{eq:sigma2}) are transformed into
\begin{eqnarray}
\sigma^{\rm I}_{\rm AH}&&=\frac{1}{2\pi N}\sum_{{\boldsymbol k},l\neq m}
j^{ml}_X j^{lm}_Y\frac{1}{(\mu-E^l_{\boldsymbol k}+i\gamma)(\mu-E^m_{\boldsymbol k}-i\gamma)}, \label{eq:sigmaI}\\
\sigma^{\rm II}_{\rm AH}&&=\frac{i}{2\pi N}\sum_{{\boldsymbol k},l\neq m}\int_{-\infty}^\mu d\epsilon j^{ml}_X j^{lm}_Y\nonumber\\
&&{\rm Im}\Biggl[ \frac{1}{(\epsilon-E^l_{\boldsymbol k}+i\gamma)^2(\epsilon-E^m_{\boldsymbol k}+i\gamma)}\nonumber\\
&&-\frac{1}{(\epsilon-E^l_{\boldsymbol k}+i\gamma)(\epsilon-E^m_{\boldsymbol k}+i\gamma)^2}\Biggr].\label{eq:sigmaII}
\end{eqnarray}
at zero temperature. Here, $l$ and $m$ are the band indices, and we dropped the diagonal terms $l=m$
since their contribution vanishes identically. 
We perform the numerical calculation for the AHC using Eqs. (\ref{eq:sigmaI}) and (\ref{eq:sigmaII}) in later section.

$\sigma^{\rm I}_{\rm AH}$ and $\sigma^{\rm II}_{\rm AH}$ are called the Fermi surface term and the Fermi sea term, respectively.
According to Refs. \cite{TanakaKontani,Streda},
$\sigma^{\rm II}_{\rm AH}$ can be uniquely divided into 
$\sigma^{\rm IIa}_{\rm AH}$ and the Berry curvature term 
$\sigma^{\rm IIb}_{\rm AH}$.
The intrinsic AHC is given by $\sigma^{\rm IIb}_{\rm AH}$ when $\gamma\rightarrow 0$ since $\sigma^{\rm I}_{\rm AH}+\sigma^{\rm IIa}_{\rm AH}=0$. 
In general cases, however, the total AHC is not simply given by 
$\sigma_{\rm AH}^{\rm IIb}$ since $\sigma^{\rm I}_{\rm AH}+\sigma^{\rm IIa}_{\rm AH}$
is finite when $\gamma\ne0$ or $\gamma_l/\gamma_m\ne1$. 
Therefore, we calculate the total AHC 
$\sigma_{\rm AH}=\sigma^{\rm I}_{\rm AH}+\sigma^{\rm II}_{\rm AH}$
in this paper.

\section{Orbital Aharonov-Bohm effect}
\begin{figure}
\includegraphics[scale=0.4]{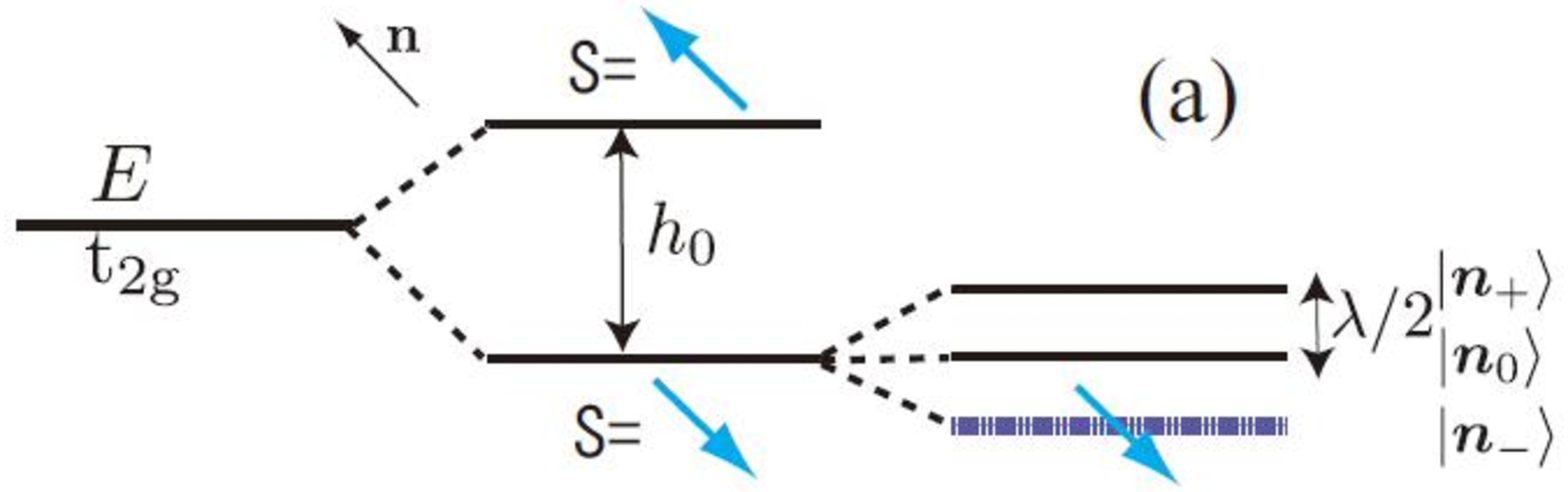}
\caption{\label{fig:energy} Eigenenergies for $t_{2g}$ electron under the exchange field ${\boldsymbol h}$
and the SOI $(-\lambda/2){\boldsymbol n}\cdot{\boldsymbol l}$;
see Ref. \cite{TomizawaKontani}.
}
\includegraphics[scale=0.2]{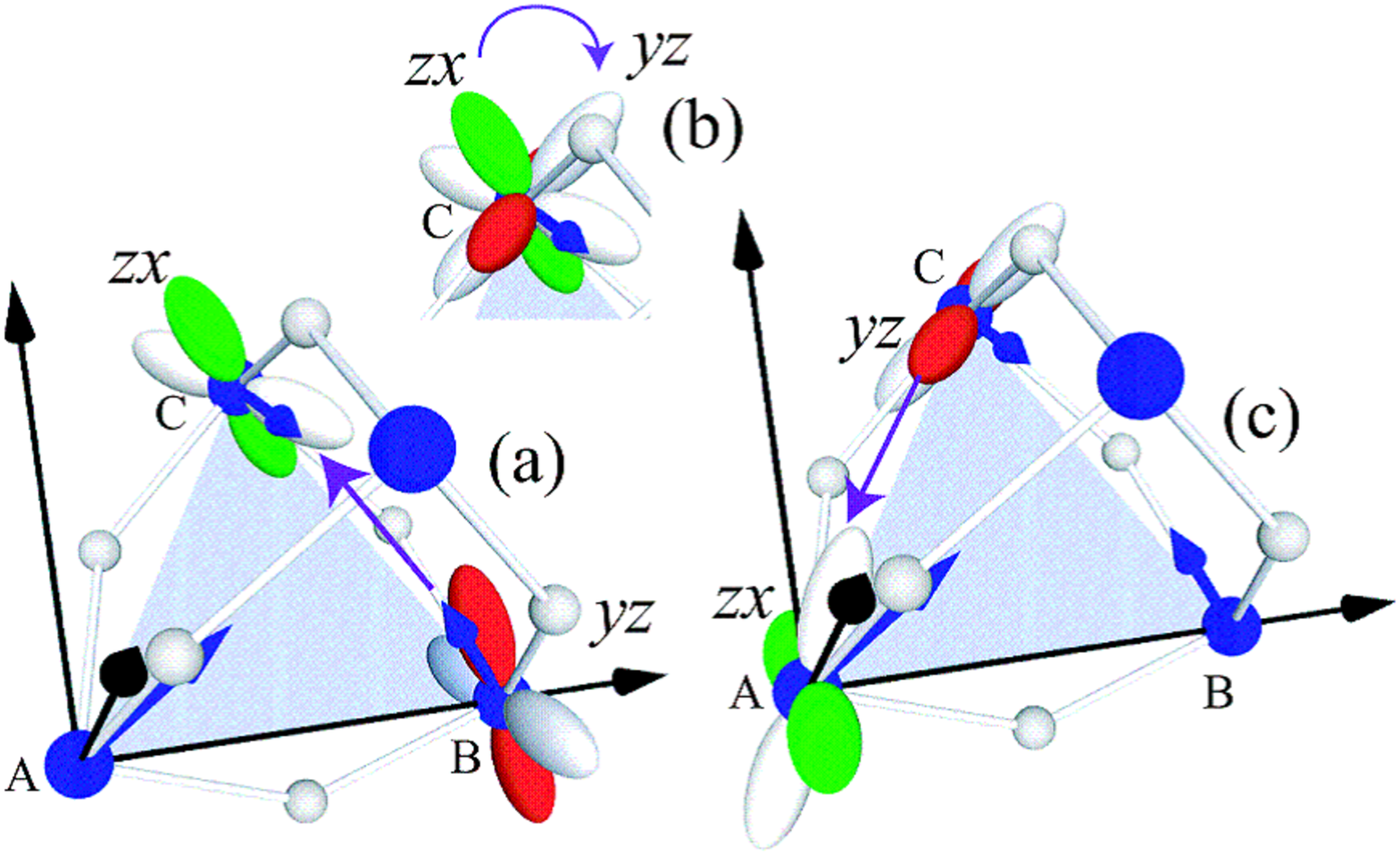}
\caption{\label{fig:ABE} Orbital AB phase given by the complex $t_{2g}$ orbital wavefunction at site C. 
The electron acquires the phase difference between $zx$- and $yz$-orbitals
(orbital AB phase) via the movement (a)$\rightarrow$(b)$\rightarrow$(c).}
\end{figure}

Before proceeding to the numerical calculation for the AHE, 
we present an intuitive explanation for the unconventional AHE
induced by the non-collinear local exchange field ${\boldsymbol h}_i$.
For this purpose, we assume the strong coupling limit where the
Zeeman energy is much larger than the kinetic energy and the SOI
 \cite{TomizawaKontani}.
The $t_{2g}$ energy levels are split into the two triply-degenerate states by the Zeeman effect, as shown in Fig. \ref{fig:energy}. 
Its eigenstate for $- h_0/2$ are given by
\begin{eqnarray}
|\alpha\rangle =\sin\frac{\theta}{2}|\alpha\uparrow\rangle+ e^{i\phi}\cos\frac{\theta}{2}|\alpha\downarrow\rangle,
\label{eq:Zeeman}
\end{eqnarray}
where $\alpha=xy,yz,zx$.
In addition, we assume the SOI is much larger than the kinetic energy. 
Since ${\boldsymbol \mu}_e=-2{\boldsymbol s}$, the SOI term at site $i$ 
is replaced with $(-\lambda/2){\boldsymbol l}\cdot{\boldsymbol n}_i$, 
where ${\boldsymbol n }_i \equiv {\boldsymbol h }_i/|{\boldsymbol h }_i|$. 
Its eigenenergies in the $t_{2g}$ space are $0$ and $\pm \lambda/2$, as shown in Fig. \ref{fig:energy}. 
The corresponding eigenstates are given by \cite{TomizawaKontani}
\begin{subequations}
\begin{eqnarray}
|{\boldsymbol n}_0\rangle &=& n_z|xy\rangle+n_x|yz\rangle+n_y|zx\rangle,\\
|{\boldsymbol n}_\pm\rangle&=&\frac{1}{\sqrt{2(n_y^2+n_z^2)}}[-(n_xn_z\pm in_y)|xy\rangle\nonumber\\
&+&(n_y^2+n_z^2)|yz\rangle-(n_xn_y\mp in_z)|zx\rangle],\label{eq:npm}
\end{eqnarray}
\end{subequations}
where ${\boldsymbol n }=(n_x,n_y,n_z)_\Xi$ in the $(xyz)_\Xi$-coordinate is given by 
$[\sin\theta\cos\phi, \sin\theta\cos\phi,\cos\theta]\hat{O_\Xi}$. 
In the complex wavefunction $|{\boldsymbol n}_{-} \rangle$,
the phase of each $d$-orbital within the $\theta$-linear term 
is given in Table II.

\begin{table}
\caption{\label{tab:}Phases for $t_{2g}$ orbitals }
\begin{ruledtabular}
\begin{tabular}{c|ccc}
&$\psi_{xy}^\Xi$&$\psi_{yz}^\Xi$&$\psi_{zx}^\Xi$\\
\hline
$\Xi$=A,B,C&$\Phi^0_{xy}+\frac{3}{26}\sqrt{\frac{3}{2}}\theta$ &0 &$\Phi^0_{zx}+\frac{3}{13}\sqrt{\frac{3}{2}}\theta$\\
\end{tabular}
\end{ruledtabular}
\end{table}

Here, we explain that 
the $\theta$-dependence of the 
$d$-orbital wavefunction $|{\boldsymbol n}_{-} \rangle$
gives rise to a prominent spin structure-driven AHE
\cite{TomizawaKontani}.
Figure \ref{fig:ABE} shows the motion of an electron: 
(a) moving from $|{\rm B};yz\rangle$ to $|{\rm C};zx\rangle$,
(b) transferring form $|{\rm C};zx\rangle$ to $|{\rm C};yz\rangle$ 
at the same site, and 
(c) moving from $|{\rm C};yz\rangle$ to $|{\rm A};zx\rangle$. 
Here, we assume that the electron is in the eigenstate 
$|{\boldsymbol n}_{-} \rangle$ at each site.
The total orbital phase factor for the triangle path along 
$\rm A\rightarrow B\rightarrow C\rightarrow A$ 
is given by the phase of the following amplitude:
\begin{eqnarray}
T_{\rm orb}&=&\langle {\rm A};{\boldsymbol n}_-|\hat{H}^t| {\rm C};{\boldsymbol n}_-\rangle 
\langle {\rm C};{\boldsymbol n}_-|\hat{H}^t| {\rm B};{\boldsymbol n}_-\rangle 
 \nonumber \\
& &\times\langle {\rm B};{\boldsymbol n}_-|\hat{H}^t| {\rm A};{\boldsymbol n}_-\rangle,
\end{eqnarray}
where $\hat{H}^t$ is the kinetic term in the Hamiltonian. 
For simplicity, we take only the following largest hopping
$t=\langle {\rm A}; zx |\hat{H}^t|{\rm B}; yz \rangle
=\langle {\rm B}; zx |\hat{H}^t|{\rm C}; yz \rangle
=\langle {\rm C};zx |\hat{H}^t|{\rm A}; yz \rangle
$,
and assume that $t$ is real.
Considering that 
$|\Xi;{\boldsymbol n}_-\rangle\langle \Xi;{\boldsymbol n}_-|
\ni |\Xi;yz\rangle\langle \Xi;zx|\frac1{3}
\exp(-i(\Phi_{zx}^0+\frac{3}{13}\sqrt{\frac{3}{2}}\theta))$
for $\Xi$=A, B, and C, the hopping amplitude is expressed as
\begin{equation}
T_{\rm orb} \sim |T_{\rm orb}|e^{-i2\pi \Phi_{\rm orb}/\Phi_0},
\end{equation}
where $\Phi_0=2\pi\hbar /|e|$ is the flux quantum, and 
$2\pi\Phi_{\rm orb}/\Phi_0=3\Phi^0_{zx}+(9/13)\sqrt{3/2}\theta$
is ``the effective AB phase'' induced by the 
complex $d$-orbital wavefunction.
The large $\theta$-linear term in $\Phi_{\rm orb}$
gives rise to the large spin structure-driven AHE in $\rm Nd_2Mo_2O_7$.

Note that $\langle {\rm B};{\bm n}_-|\hat{H}^t|{\rm A};{\bm n}'_- \rangle$ 
is not actually a real number if ${\bm n}\ne{\bm n}'$,
since the rotation of the spin axis induces the phase factor;
see Eq. (\ref{eq:Zeeman}). 
This fact gives rise to the effective flux due to the spin rotation
$\Phi_{\rm spin}$; $-4\pi\Phi_{\rm spin}$ is given by the solid angle subtended by ${\boldsymbol n}_{\rm A}$, ${\boldsymbol n}_{\rm B}$
and ${\boldsymbol n}_{\rm C}$ \cite{Ohgushi}.
Thus, the total flux is given by $\Phi_{\rm tot}=\Phi_{\rm orb}+\Phi_{\rm spin}$. 
However, $\Phi_{\rm spin}\propto \theta^2$ is negligible for $|\theta|\ll1$
 \cite{TomizawaKontani}.
Since all the upward and downward ABC triangles in the kagome lattice 
are penetrated by $\Phi_{\rm tot}\approx \Phi_{\rm orb}$, 
the orbital AB effect induces prominent spin structure-driven AHE 
in $\rm Nd_2Mo_2O_7$.

\section{Numerical Study}
In this section, we perform numerical calculation for the AHC using 
Eqs. (\ref{eq:sigmaI}) and (\ref{eq:sigmaII}),
using realistic model parameters.
We use two SK parameters between the nearest neighbor Mo sites as 
${\rm SK}(-1.0,0.6,-0.1)$ and ${\rm SK}(-1.0,0.4,-0.1)$ where we represent the set of SK parameters as ${\rm SK}((dd\sigma),(dd\pi),(dd\delta))$. 
Hereafter, we put the unit of energy $|(dd\sigma)|=1$, which 
corresponds to 2000K in real compound. 
The spin-orbit coupling constant for Mo $4d$ is $\lambda=0.5$ 
\cite{TanakaKontani}. 
The number of electrons per unit cell is $N=6$ (1/3-filling) 
for $\rm Nd_2Mo_2O_7$ since the valence of Mo ion is $4+$. 
We choose $|{\boldsymbol h}_i|$ to reproduce the magnetization of Mo ion 
$1.3\mu_{\rm B}$ in $\rm Nd_2Mo_2O_7$ \cite{Yasui}. 

\begin{figure}
\includegraphics[scale=0.35]{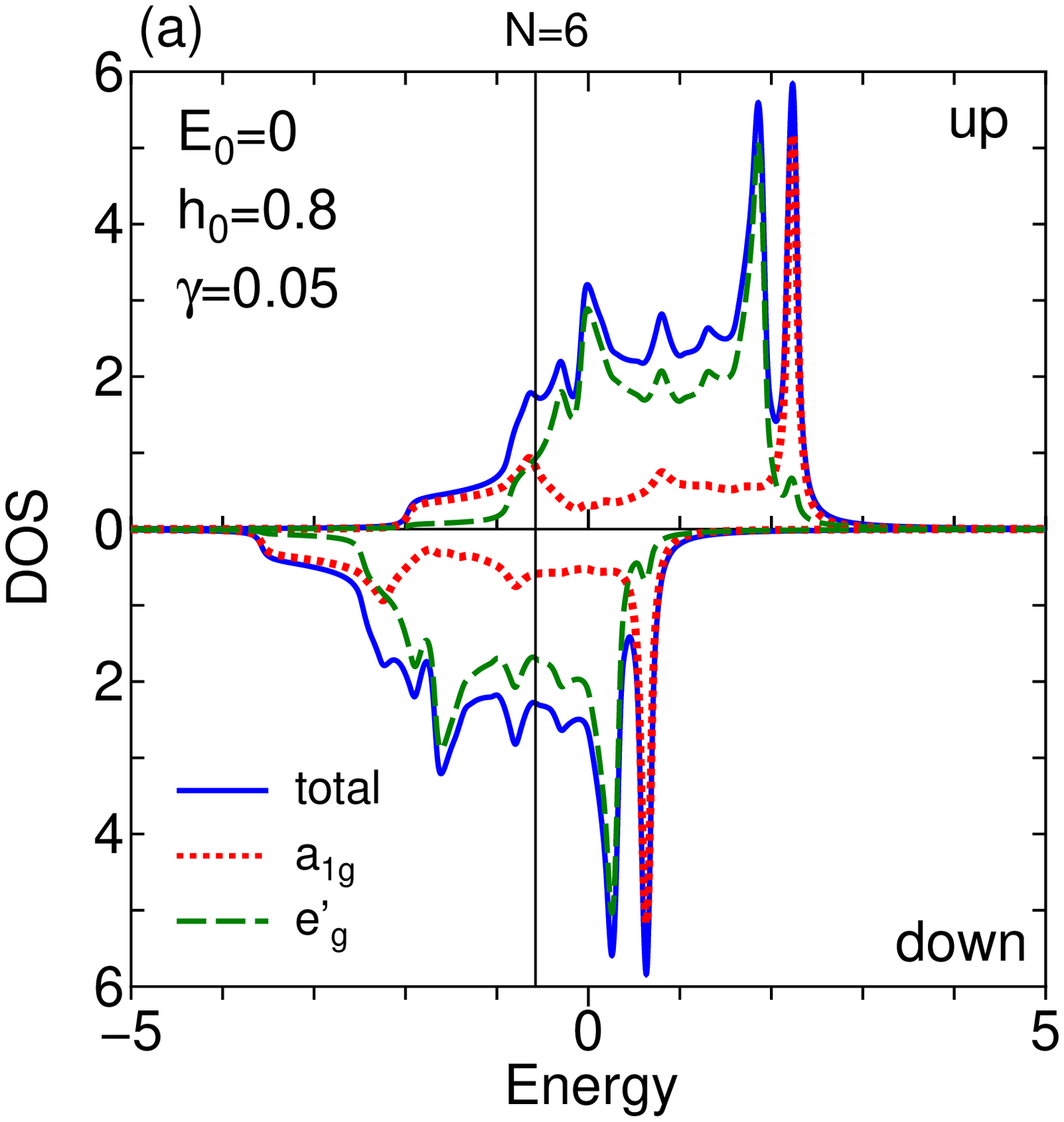}

\includegraphics[scale=0.35]{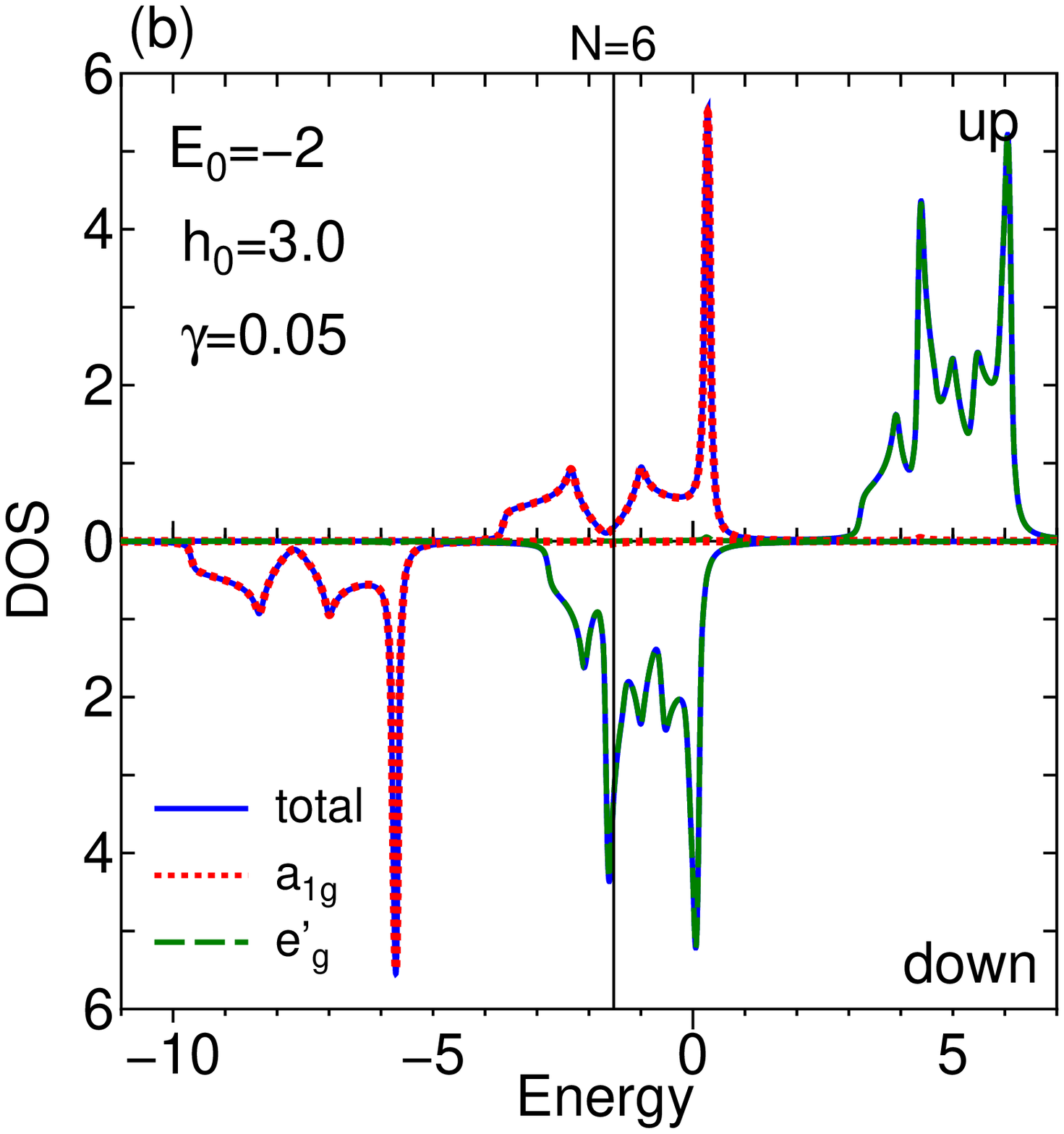}
\caption{\label{fig:DOS} Total and partial DOS for (a) $E_0=0$ and (b) $E_0=-2$}
\end{figure}

Figure \ref{fig:DOS} shows the total and partial density of states (DOS)
for ${\rm SK}(-1.0,0.6,-0.1)$ at (a) $E_0=0$ and (b) $E_0=-2$, 
with the damping rate $\gamma=0.05$.
For $E_0=0$ ($E_0=-2$)
we set $|{\boldsymbol h}_i|=0.8$ ($|{\boldsymbol h}_i|=3.0$)
to reproduce the magnetization of Mo ion $1.3\mu_{\rm B}$ \cite{Yasui}.
The crystalline electric field splitting for $E_0=-2$ in Fig. \ref{fig:DOS} (b) 
corresponds to $3|E_0|\sim 1{\rm eV}$, consistently with the 
band calculation \cite{Solovyev}. 
In both cases, the states $|a_{1g}\downarrow\rangle$ and 
$|e'_{g}\uparrow\rangle$ gives large partial DOS near the Fermi level.

\begin{figure}
\includegraphics[scale=0.3]{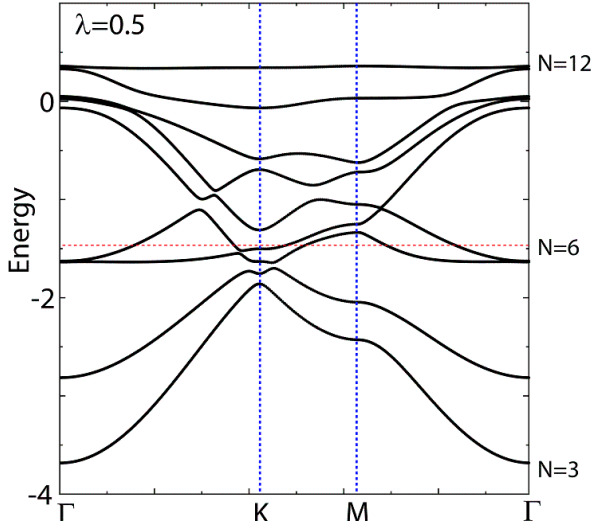}
\caption{\label{fig:disp}3rd-12th bands from lowest for $\rm SK(-1.0, 0.6, -0.1)$ and $\theta=0$.}
\end{figure}
\begin{figure}
\includegraphics[scale=0.4]{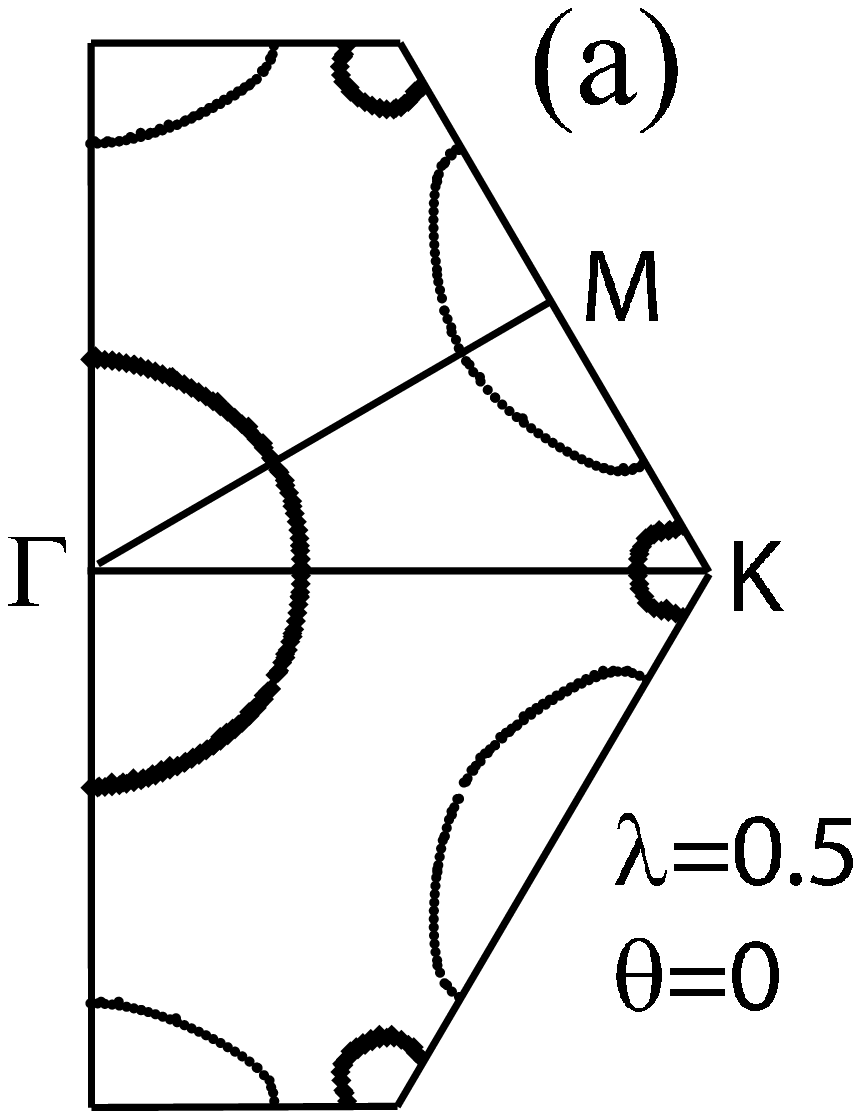}
\includegraphics[scale=0.4]{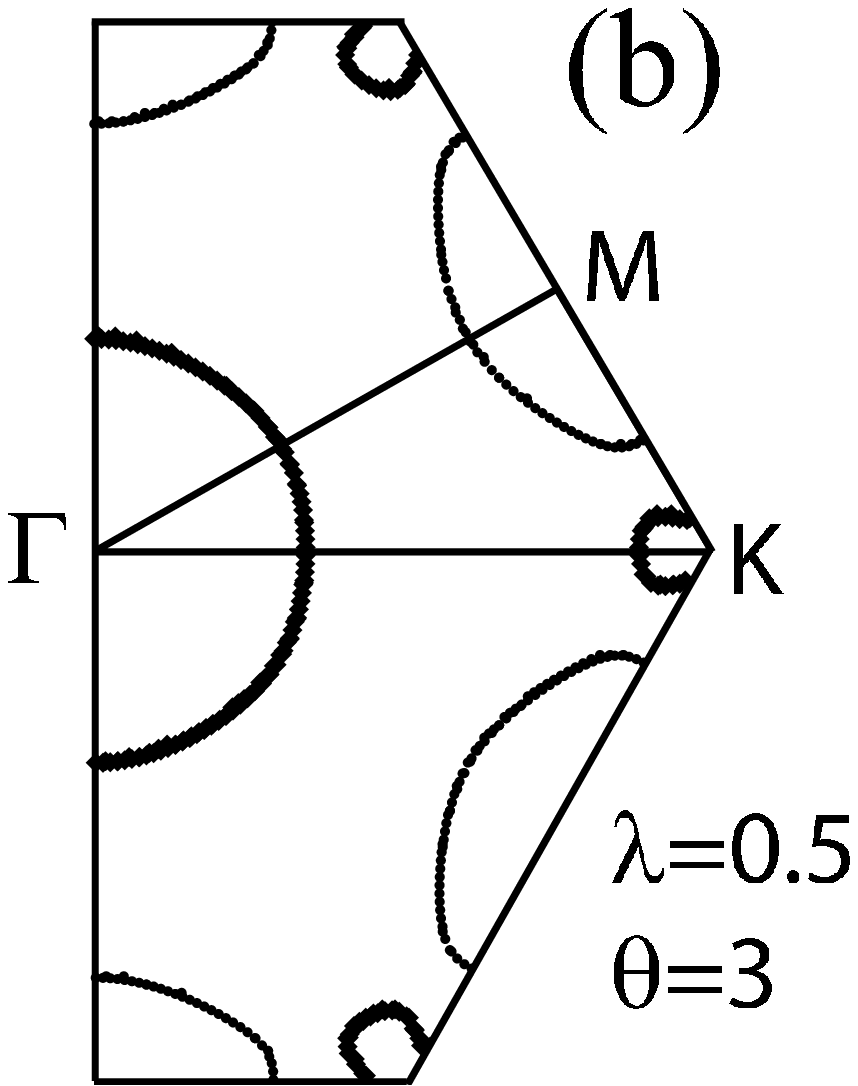}
\caption{\label{fig:FS}Fermi surface for $\rm SK(-1.0, 0.6, -0.1)$, (a) $\theta=0$, and (b) $\theta=3^\circ$.} 
\end{figure}

Figure \ref{fig:disp} shows the 3rd-12th bands from the lowest.
Nine bands near the Fermi level($N=6$) are composed of 
$|a_{1g}\downarrow\rangle$ and $|e_g'\uparrow\rangle$ 
as understood from Fig. \ref{fig:DOS}(b).
As shown in Fig. \ref{fig:FS},
the band structure and the Fermi surface are hardly changed 
by varying $\theta$ by 3 degrees.

Here, we present the numerical results of the AHC
for two SK parameter sets; 
${\rm SK}(-1.0,0.6,-0.1)$ and ${\rm SK}(-1.0,0.4,-0.1)$. 
We set $(E_0, |{\boldsymbol h}_i|)=(-2, 3.0)$ or $(0, 0.8)$;
each parameter set reproduces the magnetization of Mo ion 
$1.3\mu_{\rm B}$ \cite{Yasui}.
We also put the damping rate $\gamma=0.001$ (clean limit)
unless otherwise noted.
Hereafter, the unit of the conductivity is $e^2/ha$, 
where $h$ is the Plank constant and $a$ is the lattice constant. 
If we assume $a=4$\AA, $e^2/ha \approx 10^3\rm \Omega^{-1}cm^{-1}$.
In the numerical study, we use $512^2$ $\bm k$-meshes.

\begin{figure}
\includegraphics[scale=0.4]{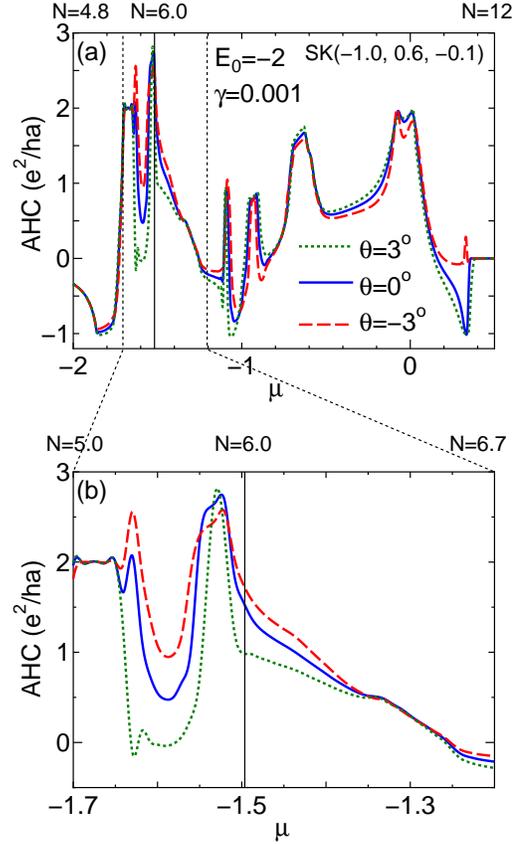}
\caption{\label{fig:mu.SK06} $\mu$-dependence of AHC for $\rm SK(-1.0, 0.6, -0.1)$, for (a) $N=4.8-12$ and (b) $N=5.0-6.7$
.}
\end{figure}

\begin{figure}
\includegraphics[scale=0.4]{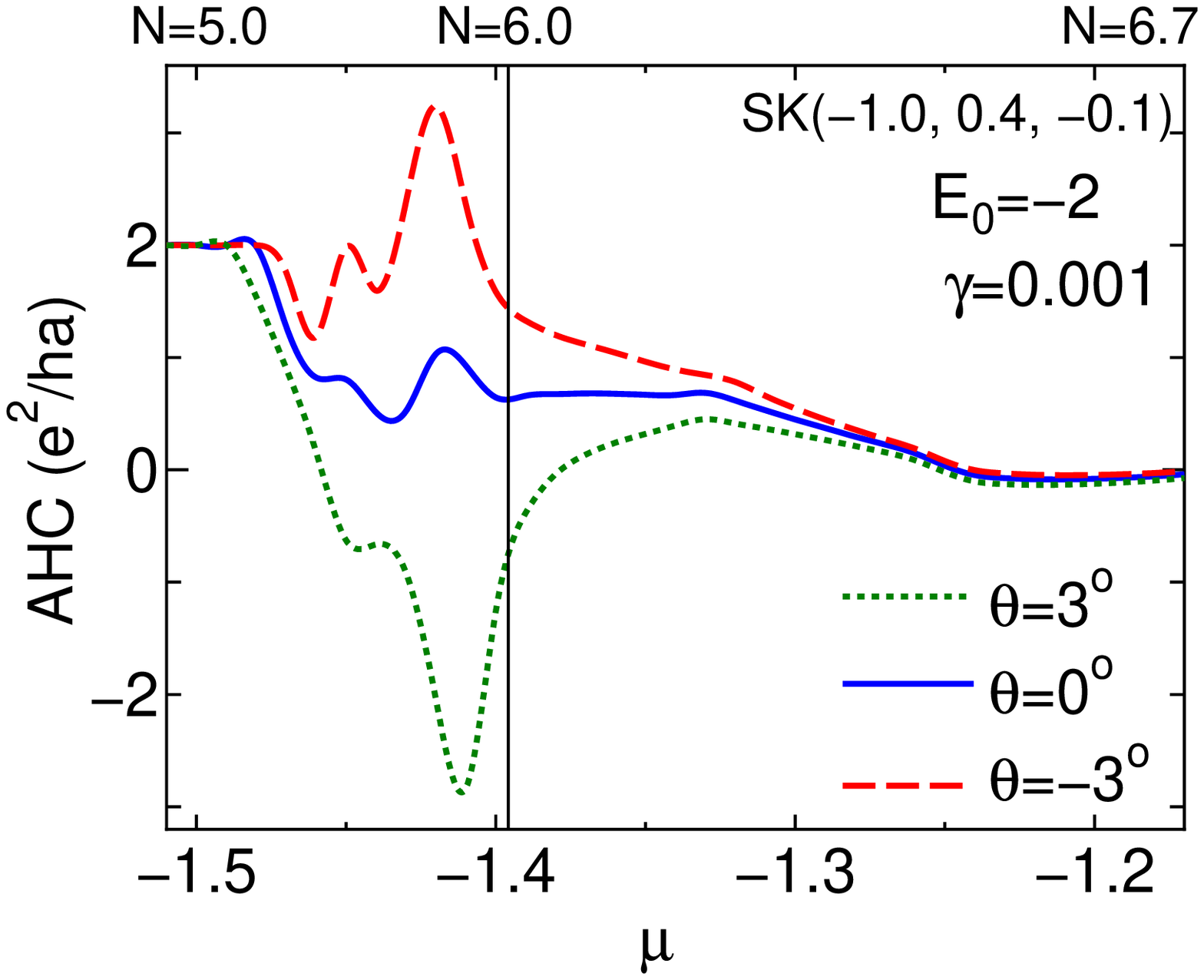}
\caption{\label{fig:mu.SK04}  $\mu$-dependence of AHC for $\rm SK(-1.0,0.4,-0.1)$ and $N=4.8-6.6$
.}
\end{figure}

Figure \ref{fig:mu.SK06} shows the obtained AHC for 
$\rm SK(-1.0,0.6,-0.1)$ at $\theta=0$ and  $\pm 3^\circ$,
for (a) a wide range of $\mu$ $(N=4.8-12)$ and 
(b) a narrow range of $\mu$ $(N=5.0-6.7)$. 
Since the present spin structure-driven AHE is linear in $\theta$, 
a very small $\theta$ causes a prominent change in the AHC 
although the Fermi surfaces are hardly changed (see Fig. \ref{fig:FS}). 
The $\mu$-dependence of the AHC for other SK parameter $\rm SK(-1.0,0.4,-0.1)$
is shown by Fig. \ref{fig:mu.SK04} for $N=4.8-6.6$. 
A remarkable change of the AHC is also caused by small change in $\theta$.
Therefore, large $\theta$-linear term in the AHC is obtained 
by using general SK parameters.

The finite AHC at $\theta=0$ is nothing but the conventional KL-term.
However, obtained $\theta$-linear AHC deviates from the conventional KL-term
that is proportional to the magnetization $M_z\propto\theta^2$.
We stress that the large $\theta$-linear term in 
Figs. \ref{fig:mu.SK06} and \ref{fig:mu.SK04} 
cannot be simply understood as the movement of Dirac points
(or band crossing points) across the Fermi level,
since the change in the band structure by $\theta=\pm3^\circ$
is very tiny as recognized in Fig. \ref{fig:FS}.
Thus, the origin of the $\theta$-linear term should be ascribed to 
the orbital AB phase \cite{TomizawaKontani} discussed in Sec. IV.

\begin{figure}
\includegraphics[scale=0.4]{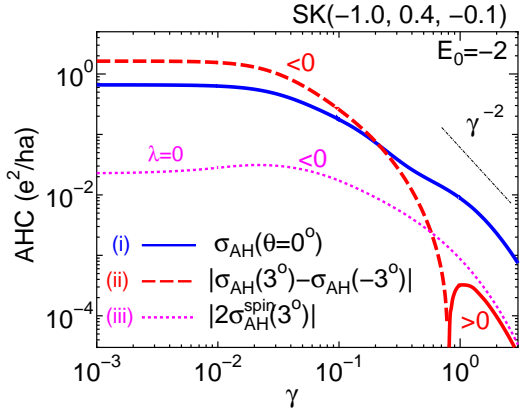}
\caption{\label{fig:gamma} $\gamma$-dependence of AHCs
for $\rm SK(-1.0,0.4,-0.1)$ and $N=6$.
The AHC starts to decrease for $\gamma\sim0.02$,
which corresponds to $\rho=0.26 {\rm m\Omega} {\rm cm}$ 
in the present parameters.
}
\end{figure}

Next, we discuss the $\gamma$-dependence of the AHC. 
As $\gamma$ increases from $0.001$, spike-like fine structure
in Figs. \ref{fig:mu.SK06} and \ref{fig:mu.SK04}
becomes moderate as recognized in Ref. \cite{TomizawaKontani}.
Moreover, the intrinsic AHC shows a crossover behavior,
that is, the AHC starts to decrease when $\gamma$ exceeds the 
band-splitting $\Delta$,
proved by using tight-binding models
 \cite{KontaniYamada,KontaniTanakaYamada,Streda2}
or local orbitals approach
 \cite{Streda2}.
Figure \ref{fig:gamma} shows the $\gamma$-dependence of the AHC 
in the present model.
Line (i) represents the total AHC for $\theta=0$; $\sigma_{\rm AH}(\theta=0)$, 
and line (ii) represents the variation of the AHC from $\theta=-3^\circ$ to 
$3^\circ$; $|\sigma_{\rm AH}(3^\circ)-\sigma_{\rm AH}(-3^\circ)|$. 
We also calculate the AHC for $\lambda=0$,
which represents the spin chirality driven AHC $\sigma_{\rm AH}^{\rm spin}$.
Note that $\sigma_{\rm AH}^{\rm spin}(\theta)$ is an even function of $\theta$,
and $\sigma_{\rm AH}^{\rm spin}(0)$=0. 
In Fig. \ref{fig:gamma}, we plot $|2\sigma_{\rm AH}^{\rm spin}(3^\circ)|$ 
as line (iii).
The variation of the AHC from $\theta=-3^\circ$ to $3^\circ$ due to 
the orbital mechanism is 100 times larger 
than the spin chirality term in the clean limit. 
Note that the intrinsic AHC follows an approximate scaling relation $\sigma_{\rm AH}\propto \rho^2$ \cite{KontaniYamada,KontaniTanakaYamada,Streda2} 
in the ``high-resistivity regime". 
In Fig. \ref{fig:gamma}, we see that $|\sigma_{\rm AH}(3^\circ)-\sigma_{\rm AH}(-3^\circ)|$ also follows the relation $\rho^2$ similarly. 

\begin{figure}
\includegraphics[scale=0.4]{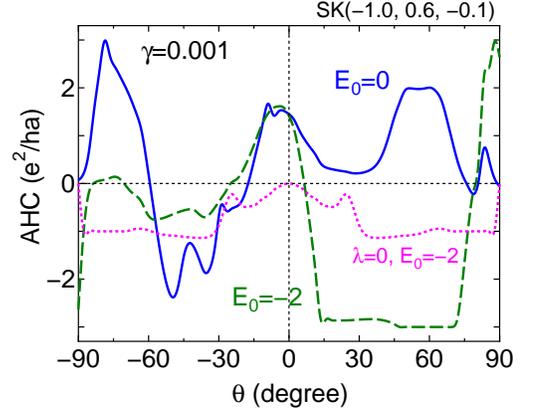}
\caption{\label{fig:theta} $\theta$-dependence of AHCs 
for $\rm SK(-1.0,0.6,-0.1)$, $N=6$ and $\gamma=0.001$,
in cases of $(E_0, \lambda)=(-2,0.5)$, $(E_0, \lambda)=(0,0.5)$, 
and $(E_0, \lambda)=(-2,0)$.
}
\end{figure}

\begin{figure}
\includegraphics[scale=0.4]{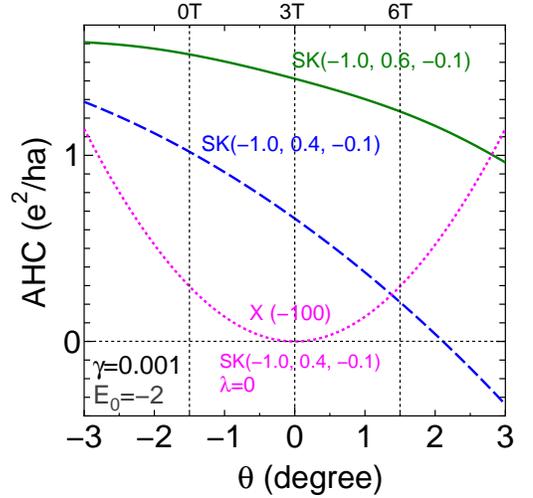}
\caption{\label{fig:theta2}  $\theta$-dependence of 
AHCs for $|\theta|\leq 3^\circ$ at $\gamma=0.001$.
The spin chirality term ($\lambda=0$) is very small.
}

\end{figure}

Next, we analyze the overall $\theta$-dependence of the AHC, by ignoring the experimental condition $|\theta|\ll1$. 
Figure \ref{fig:theta} shows the AHCs as functions of $\theta$. 
Solid and dashed lines represent the AHCs for $E_0=0$ and $-2$, respectively. 
They have large $\theta$-linear terms for $\theta\sim0$, 
and they take finite values even if $\theta=\pm\pi/2$ (coplanar order). 
Note that obtained $\theta$-dependence of the AHC 
is insensitive to the value of $E_0$.
The AHCs for $\theta=0$ corresponds to the conventional KL-type AHE.
Dotted line in Fig. \ref{fig:theta} shows the AHC for $\lambda=0$, 
which gives the spin chirality term $\sigma_{\rm AH}^{\rm spin}$.
It is proportional to $\theta^2$ for small $\theta$, 
and becomes zero when $\theta=\pm\pi/2$. 
Finally, we analyze the $\theta$-dependence of the AHC more in detail 
for $|\theta|\leq 3^\circ$ in Fig. \ref{fig:theta2}. 
In the case of $\rm SK(-1,0.4,-0.1)$, 
the AHC for $\lambda=0.5$ changes the sign at $\theta\sim 2^\circ$
due to the orbital AB effect,
and it is more that 100 times larger than the AHC for 
the spin chirality term ($\lambda=0$).




\section{\label{section: Pr2Ir2O7} $\rm Pr_2 Ir_2 O_7$}
In the previous section, we discussed the unconventional AHE 
in the pyrochlore $\rm Nd_2Mo_2O_7$. 
Here, we discuss other pyrochlore $\rm Pr_2Ir_2O_7$. 
Unlike Mo $4d$ electrons in $\rm Nd_2Mo_2O_7$, Ir 5$d$ electrons are in the paramagnetic state. 
Below $\theta_{\rm W}=1.7$K, localized Pr 4$f$ electrons form non-coplanar spin-ice magnetic order. 
Under the magnetic field along [111], 
the non-coplanar structure of Pr Ising moments are expected to change from ``2in 2out"$(H\sim 0.7{\rm Tesla})$ to ``3in 1out"$(H>0.7{\rm Tesla})$. 
The AHC increases in proportion to the magnetization with field from 
0 to 0.7 Tesla, 
whereas it rapidly decreases as the spins of Pr tetrahedron 
change from ``2in 2out" to ``3in 1out" for $H>0.7$Tesla.

On Ir sites in $\rm Pr_2Ir_2O_7$, the tilted ferromagnet state shown in Fig. \ref{fig:TiltedFerro} is also realized. 
In $\rm Pr_2Ir_2O_7$, however, the ferromagnetic exchange interaction 
is absent, and the local exchange field $\bm{h}_i$ on Ir ion is 
composed of only the exchange field from the Pr moment; $\sim J_{df}$. 
Since $\bm{h}_i$ is parallel to the sum of the nearest Pr momenta, 
$\theta$ of Ir spin is much larger than the 
$\theta$ of Mo spin in $\rm Nd_2Mo_2O_7$. 
Therefore, the tilted ferromagnetic state with large $\theta$ and 
small $|\bm{h}_i|$ is realized in $\rm Pr_2Ir_2O_7$. 

\begin{figure}
\includegraphics[scale=0.23]{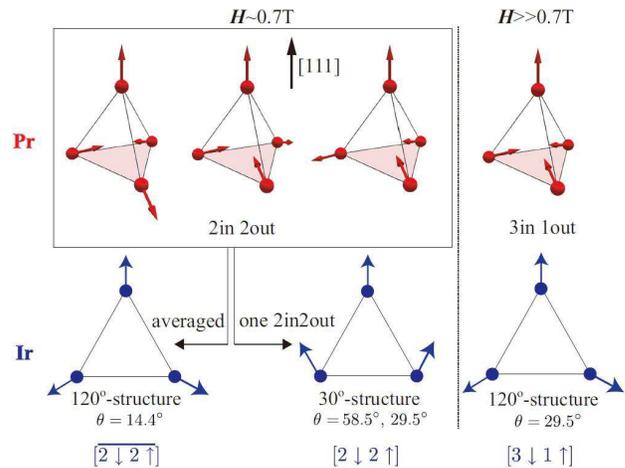}
\caption{\label{fig:Pr-structure}
The spin configurations in Pr tetrahedron 
are shown in the first line. 
Three ``2in 2out" states are realized under $H\lesssim 0.7$ Tesla, 
and one ``3in 1out" state is realized under the higher field.
These Pr spin configurations induce the local exchange fields 
at Ir sites as shown in the last line.
In the $120^\circ$-structure,
$(\phi_{\rm A},\phi_{\rm B},\phi_{\rm C})=(-5\pi/6, -\pi/6, \pi/2)$.
In the $30^\circ$-structure,
$(\phi_{\rm A},\phi_{\rm B},\phi_{\rm C})=(2\pi/3,\pi/3,\pi/2)$.
}
\label{fig:Pr}
\end{figure}

Now, we explain the local exchange field on Ir sites
given by Pr tetrahedron. 
Details of the derivation of these local exchange field are presented 
in Appendix B. 
In the strong magnetic field along [111] $(>\!\!>0.7{\rm Tesla})$, 
the spins of Pr tetrahedron have ``3in 1out" structure, 
and the realized local exchange fields at Ir sites are
$(\phi_{\rm A},\phi_{\rm B},\phi_{\rm C})=(-5\pi/6, -\pi/6, \pi/2)$
and $\theta=29.5^\circ$ in Fig. \ref{fig:Pr-structure}.
We denote this Ir spin structure as $[3\downarrow1\uparrow]$. 
In this section, we promise that $0\le\theta\le\pi$ and $-\pi\le\phi\le\pi$.
In the intermediate field ($\sim 0.7{\rm Tesla}$), the spin of Pr tetrahedron 
can take three types of ``2in 2out" structures with negative Zeeman energy.
If we take one ``2in 2out" structure among three, the exchange fields 
at Ir sites are $(\phi_{\rm A},\phi_{\rm B},\phi_{\rm C})=(2\pi/3,\pi/3,\pi/2)$,
$\theta_{A,B}=58.5^\circ$, and $\theta_C=29.5^\circ$ in Fig. \ref{fig:Pr-structure}.
We denote this Ir spin structure as $[2\downarrow2\uparrow]$.
In real compounds, domain structures of three ``2in 2out" structures
are expected to be formed, and the total magnetization is parallel to $Z$-axis.
The total AHC will be insensitive to the domain structure 
since $\sigma_{\rm AH}$'s due to three $[2\downarrow2\uparrow]$ structures
are equivalent.
If we take average of three ``2in 2out" structures, 
the local exchange fields belongs to the $120^\circ$-structure
with $\theta=14.4^\circ$, as shown in Fig. \ref{fig:Pr-structure}.
We denote this Ir spin structure as $[\overline{2\downarrow2\uparrow}]$.
As a result, the Ir spin structure changes as 
$[2\downarrow2\uparrow]$ (or $[\overline{2\downarrow2\uparrow}]$)
$\rightarrow$$[3\downarrow1\uparrow]$ 
with increasing the field from $\sim$0.7 Tesla gradually. 

\begin{figure}
\includegraphics[scale=0.4]{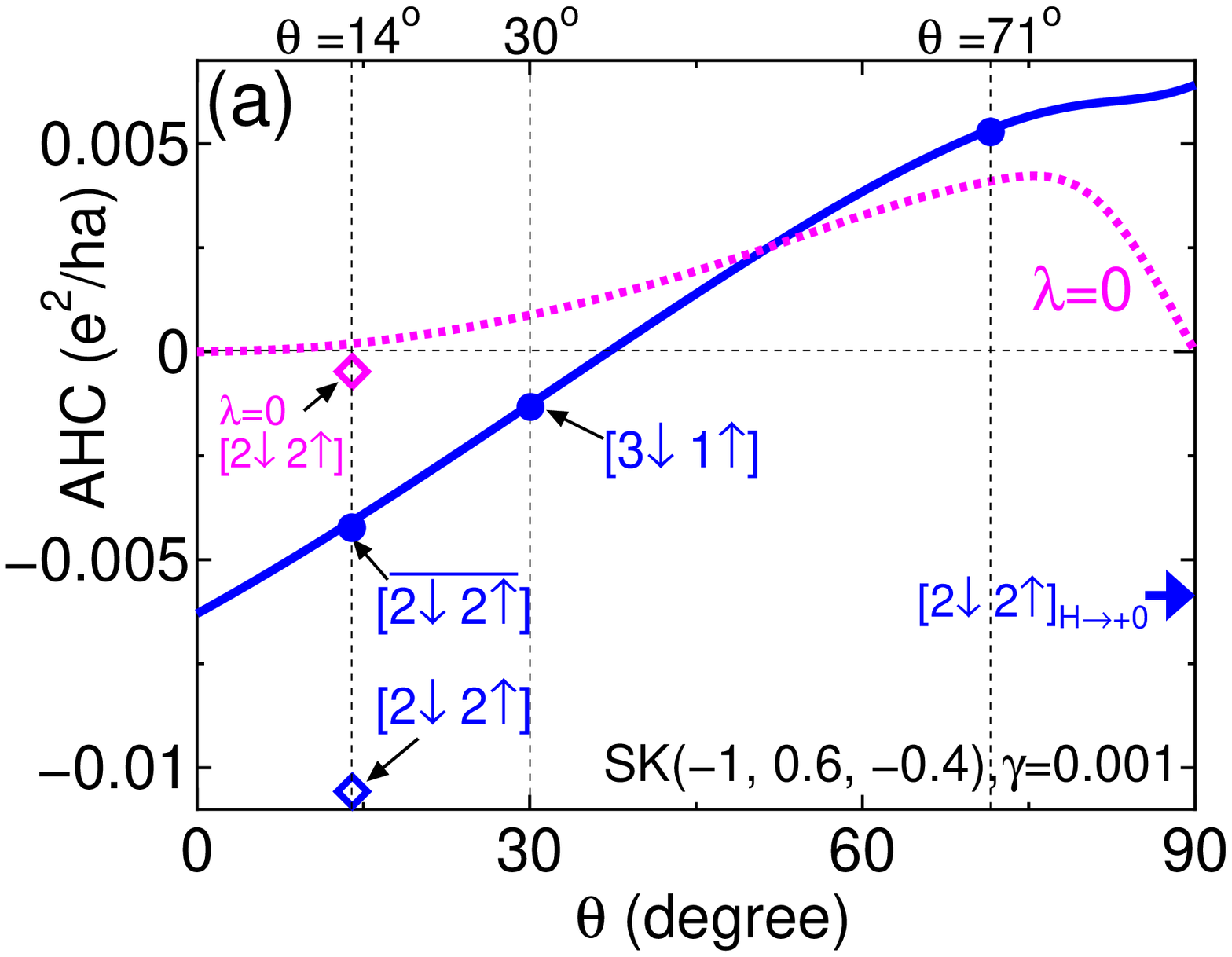}
\includegraphics[scale=0.4]{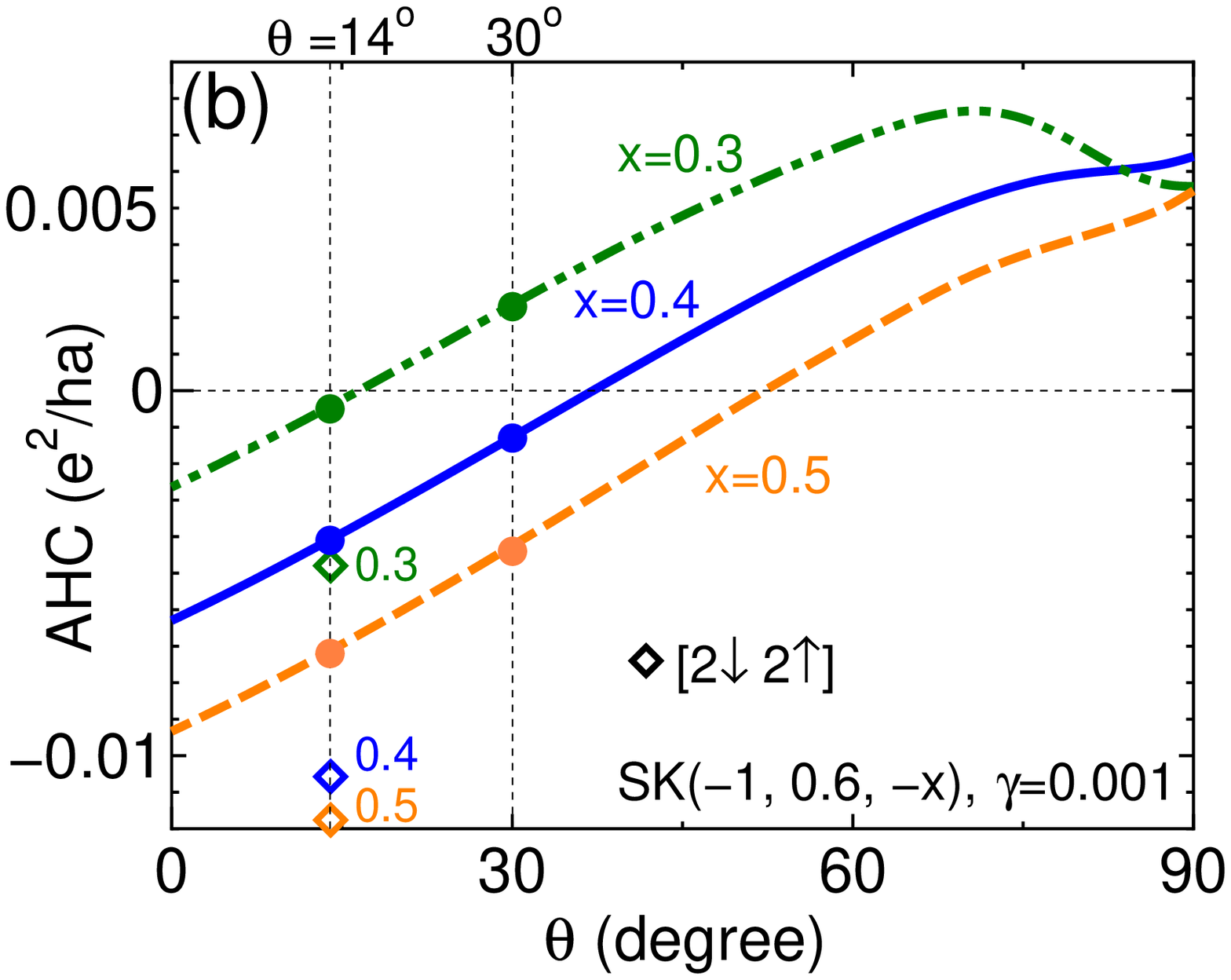}
\includegraphics[scale=0.35]{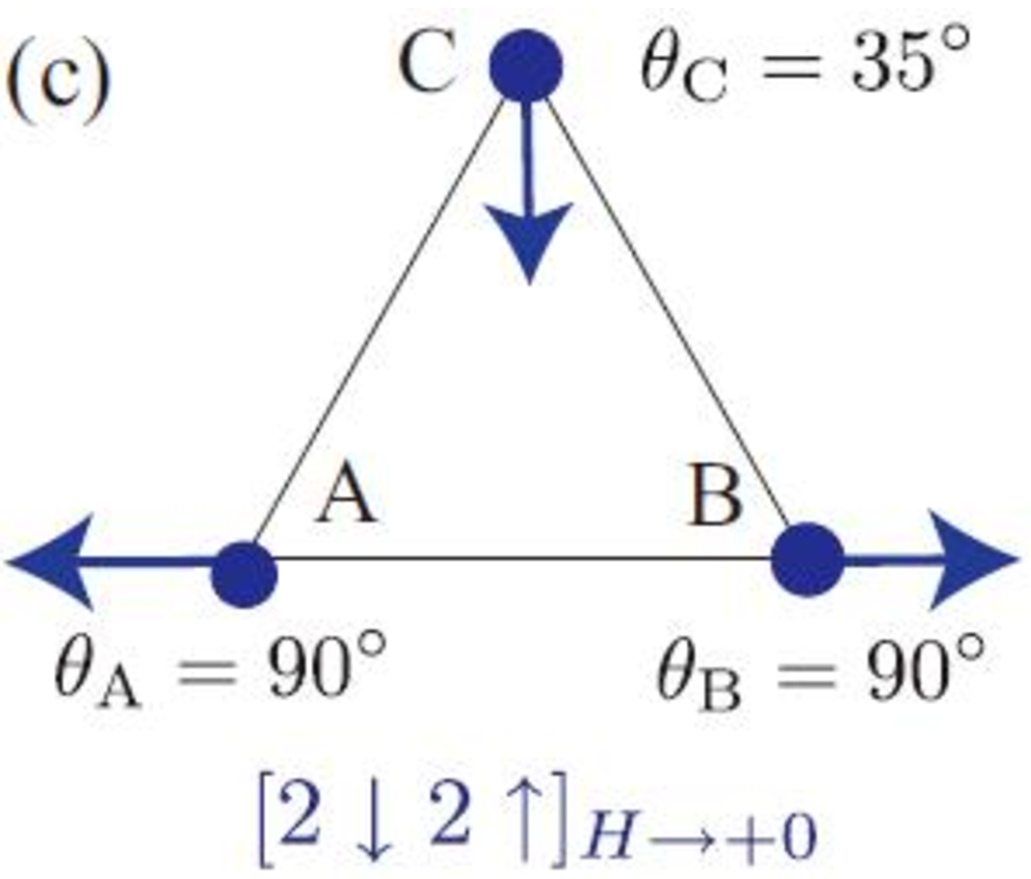}
\caption{\label{fig:thetaPrIr}
$\theta$-dependence of AHC in $\rm Pr_2Ir_2O_7$ 
for (a) ${\rm SK}(-1.0,0.6,-0.4)$ and (b) ${\rm SK}(-1.0,0.6,-x)$ 
where $x=0.3\sim0.5$. 
Each line represents the AHC for the $120^\circ$-structure in
Fig. \ref{fig:Pr}.
(c) 
Ir spin structure below $T_{\rm N}^{\rm Ir}$ under the 
weak exchange field ${\bm h}_i$ due to ``2in 2out'' Pr spin-ice order;
$[2\downarrow2\uparrow]_{H\rightarrow+0}$.
The total magnetization of the Ir tetrahedron is zero
since ${\bm S}_{\rm C}=-{\bm S}_{\rm D} \parallel$ {\bf{CD}},
where site D is the apical Ir site.
Note that $[2\downarrow2\uparrow]_{H\rightarrow-0}$ spin structure 
is the reverse of $[2\downarrow2\uparrow]_{H\rightarrow+0}$.
}
\end{figure}

Here, we perform the numerical calculation for $\rm Pr_2Ir_2O_7$. 
We put the atomic SOI as $\lambda=3000{\rm K}$, 
which is slightly smaller than the atomic value for Ir \cite{TanakaKontani}. 
The number of electrons per unit cell is $N=15$ for $\rm Pr_2Ir_2O_7$. 
We set $|{\boldsymbol h}_i|=20{\rm K}$ since $|{\boldsymbol h}_i|\sim J_{df}$ 
is estimated to be larger than 14K experimentally \cite{Machida}. 
We also put the damping rate $\gamma=0.001$ (clean limit). 
Figure \ref{fig:thetaPrIr} (a) shows the AHC in 
$\rm Pr_2Ir_2O_7$ with ${\rm SK}(-1.0,0.6,-0.4)$.
Each line represents the AHC for the $120^\circ$-structure in Fig. \ref{fig:Pr}.
The line with ``$\lambda=0$'' represents the spin chirality term.
In the case of $[2\downarrow2\uparrow]$,
the AHC in the present model is 10 times larger than the AHC for $\lambda=0$.
Thus, the orbital AB effect dominates the chirality mechanism.
The variation of the AHC for
$[2\downarrow2\uparrow]$ (or $[\overline{2\downarrow2\uparrow}]$)
$\rightarrow$$[3\downarrow1\uparrow]$ 
can explain the experimental results, ignoring the sign of the AHC. 
For example, the sign of the AHC is changed if $J_{df}$ is negative.


In Fig. \ref{fig:thetaPrIr} (b),
we put ${\rm SK}(-1.0,0.6,-x)$ with $x=0.3,0.4$ and $0.5$. 
Although the KL term at $\theta=0$ decreases from negative to positive
with $x$, the overall $\theta$-dependence of the AHC is not very
sensitive to $x$.

Recently, Ref. \cite{MachidaNature} reports that 
the AHC in $\rm Pr_2Ir_2O_7$ shows a hysteresis behavior 
under the magnetic field below $T^H\approx1$K.
That is, the AHC shows the 
``residual AHE with zero magnetization'' in $\rm Pr_2Ir_2O_7$.
In terms of the spin chirality mechanism,
the authors claimed the existence of 
a long-period magnetic (or chirality) order of Pr sites 
with 12 original unit cells 
\cite{MachidaNature}. 
However, there is no theoretical justification for this complex state.
Even if it is justified, the origin of the hysteresis behavior is unclear.
In addition, the magnetic susceptibility $\chi^s\approx \chi^s_{\rm Pr}$ 
does not show anomaly at $T^H$ experimentally.

Here, we propose an alternative explanation for the residual AHE
based on the orbital AB effect:
In $Ln_2$Ir$_2$O$_7$ with $Ln$=Nd, Sm, and Eu, the Ir 5$d$-electrons 
show magnetic order at $T_{\rm N}^{\rm Ir}=$36 K, 117 K, and 120 K, respectively
\cite{Matsuhira}.
Thus, $T_{\rm N}^{\rm Ir}$ monotonically decreases as the radius of 
$Ln$ ion increases.
Since Pr is on the left-hand-side of Nd in the periodic table,
one may expect a finite $T_{\rm N}^{\rm Ir}(\sim 1 {\rm K})$ in Pr$_2$Ir$_2$O$_7$.
We stress that small amount of impurities could induce the 
magnetic order in the vicinity of magnetic quantum-critical-point
\cite{Kontani-ROP}.
Here, we analyze the Ir spin structure below $T_{\rm N}^{\rm Ir}$,
considering the classical Heisenberg model for Ir tetrahedron
under the exchange field ${\bm h}_i$ by Pr spins 
(see in Appendix B):
\begin{eqnarray}
E=J(\sum_{i=1}^4 {\bm S}_i)^2-\sum_{i=1}^4 {\bm h}_i \cdot {\bm S}_i,
\label{eqn:Heisen}
\end{eqnarray}
where ${\bm S}_i$ is the $i$-th Ir spin, and the positive 
$J\ (\sim T_{\rm N}^{\rm Ir})$ is the 
antiferromagnetic interaction between Ir spins.
When $J\ll |{\bm h}_i|$, then ${\bm S}_i$ is parallel to ${\bm h}_i$.
When $J\gg |{\bm h}_i|$, we have to find the spin configuration
to minimize eq. (\ref{eqn:Heisen})
under the constraint $\sum_{i=1}^4 {\bm S}_i=0$.

Under the exchange field by one of `2in 2out'' Pr order,
the obtained Ir spin structure for $J\gg |{\bm h}_i|$
is shown in Fig. \ref{fig:thetaPrIr}.
For $J\gg |{\bm h}_i|$ under $H=+0$ Tesla, the Ir spin structure is 
changed to the $120^\circ$-structure with $\theta=70.5^\circ$ 
in Fig. \ref{fig:Pr-structure} (c),
which we denote $[2\downarrow2\uparrow]_{H\rightarrow+0}$ structure.
The obtained AHC under this spin structure is $-0.006$,
as denoted in Fig. \ref{fig:Pr-structure} (a):
The experimental residual AHC is smaller, since the
Ir ordered moment is expected to be smaller experimentally.
The AHC is reversed under $H=-0$ Tesla since 
the Ir spin structure is reversed.
As a result, we can naturally explain the ``hysteresis behavior of the AHC''
below $T^H\sim 1$ K reported in Ref. \cite{MachidaNature}.
We stress that the spin chirality term is zero under the
$[2\downarrow2\uparrow]_{H\rightarrow\pm0}$ structure,
since ${\bm S}_A\cdot({\bm S}_B\times{\bm S}_C)=0$.

Also, the Ir spins
under the exchange field by the averaged ``2in 2out'' Pr order
for $J\gg |{\bm h}_i|$ show the $120^\circ$-structure
in Fig. \ref{fig:Pr-structure} with $\theta=70.5^\circ$:
We denote this structure as 
$[\overline{2\downarrow2\uparrow}]_{H\rightarrow+0}$.
The total magnetization is zero since the Ir spin on the apical site 
(not shown) is antiparallel to the $Z$-axis.
In this case, we can also explain the ``hysteresis behavior of the AHC''
below $T^H\sim 1$ K.
However, the sign of the AHC for 
$[\overline{2\downarrow2\uparrow}]_{H\rightarrow+0}$ is different from that for 
$[\overline{2\downarrow2\uparrow}]$ under the positive $H$.

\section{Discussion}
\subsection{Comparison between theory with experiments}
First, we compare the theory with experiments for 
$\rm Nd_2Mo_2O_7$ \cite{Yoshii, Kageyama, Yasui, Sato} in detail. 
Under ${\boldsymbol H}||[111]$ below $T_{\rm N}$, 
$\sigma_{\rm AH}$ monotonically decreases with $H$ from 
$0{\rm Tesla}$ $(\theta\approx -1.5^\circ)$ to 
$6{\rm Tesla}$ $(\theta\approx 1.5^\circ)$. 
This monotonic decreasing in AHC can be explained by the 
$\theta$-linear term in the present model. 
The relation $\rho_{\rm H}\sim 4\pi R_{\rm s}M^{\rm Mo}_Z+4\pi R'_{\rm s}M^{\rm Nd}_Z$ 
describes the experimental results well,
where $M^{\rm Mo}_Z(M^{\rm Nd}_Z)$ and $R_{\rm s}(R'_{\rm s})$ 
are the magnetization and anomalous Hall coefficients for the Mo(Nd) moment
\cite{Yoshii, Kageyama, Yasui}.
In this equation, the first term represents the conventional AHE
that is recognized as the KL mechanism.
In contrast, the second term is highly unusual in that 
Nd electrons are totally localized;
it represents the unconventional AHE 
due to the non-collinear spin configuration.
As the magnetic field increases from 0 Tesla to 6 Tesla,
$M^{\rm Nd}_Z$ increases from negative to positive.
Since $\theta\propto M^{\rm Nd}_Z$ \cite{TomizawaKontani}, 
the second term corresponds to the $\theta$-linear term 
given by the orbital AB effect.
Moreover, the AHC for $[2\downarrow2\uparrow]_{H\rightarrow+0}$ in 
Fig. \ref{fig:thetaPrIr} is finite, 
irrespective of the absence of magnetization.

Next, we compare the present theory with experiments for 
$\rm Pr_2Ir_2O_7$ \cite{Machida}. 
Under the magnetic field along [111], the observed AHC increases 
in proportion to the magnetization with field from 0 Tesla, 
whereas it decrease with $H$ above $0.7$ Tesla as the spins of Pr tetrahedron 
start to change from ``2in 2out" to ``3in 1out".
The peak value of AHC around 0.7 Tesla is 17$\rm\Omega^{-1}cm^{-1}$. 
The AHCs in Fig. \ref{fig:thetaPrIr} (a) are
-11$\rm\Omega^{-1}cm^{-1}$ for [$2\downarrow2\uparrow$],
and it is doubled if we put ${\boldsymbol h}_i\rightarrow 2{\boldsymbol h}_i$.
Thus, the variation of the AHC for
$[2\downarrow2\uparrow]$ 
$\rightarrow$$[3\downarrow1\uparrow]$ in Fig. \ref{fig:thetaPrIr} (a)
can explain the experimental field dependence. 
The obtained AHC is mainly given by the orbital AB effect,
and the spin chirality term is too small to reproduce experimental values.

\subsection{Second-order-perturbation theory for 
spin structure-driven AHCs}

\begin{figure}
\includegraphics[scale=0.3]{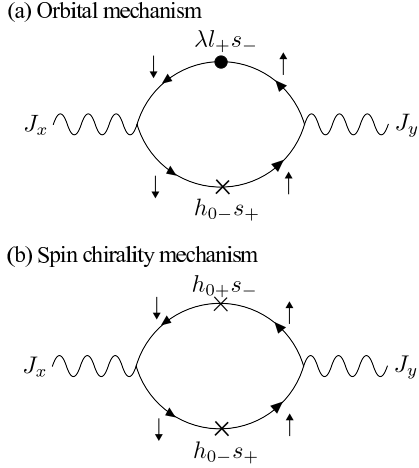}
\caption{\label{fig:sigmadiagram}
Examples of the second-order diagrams for spin structure-driven AHCs:
(a) orbital mechanism and (b) spin chirality mechanism.
The conventional AHE (KL-term) is given by the diagrams (a)
by replacing $\lambda l_+s_-$ and $h_{0-}s_+$
with $\lambda l_zs_z$ and $h_{0z}s_z$, respectively.
}
\end{figure}

Here, we discuss the spin structure-driven AHC
based on the second-order-perturbation theory
with respect to $\lambda$ and ${\bm h}_0$. 
The present weak-coupling analysis together with the strong-coupling 
analysis in Sec. IV will provide us useful complementary understanding.
Since their expressions in the present model are too complicated,
we show only some examples of the the second order diagrams 
for the spin structure-driven AHCs in Fig. \ref{fig:sigmadiagram}:
(a) $\sigma_{\rm AH}^{\rm orb}$ due to the orbital mechanism,
and (b) $\sigma_{\rm AH}^{\rm spin}$ due to the spin chirality mechanism.
In (a), the spin of conduction electron is flipped by
$x,y$-components of the Zeeman term, $h_{0\pm}s_\mp$,
and the SOI term, $\lambda l_\pm s_\mp$,
and the obtained SHC is $\sigma_{\rm AH}^{\rm orb}\sim h_{0\pm}\lambda/\Delta$, 
where $\Delta$ is the band splitting near the Fermi level. 
We stress that this term vanishes when $C_2$ rotational symmetry 
along $Z$-axis exists \cite{TomizawaKontani}:
In the present model, the $C_2$ rotational symmetry of the 
simple kagome lattice in Fig. \ref{fig:pyroXYZ} (b) is violated
by the existence of oxygen atoms.
In Fig. \ref{fig:sigmadiagram} (b), the spin is flipped by $h_{0\pm}s_\mp$ twice,
and it is given by $\sigma_{\rm AH}^{\rm spin}\sim h_{0\pm}^2/\Delta$.
Thus, $\sigma_{\rm AH}^{\rm orb}$ and $\sigma_{\rm AH}^{\rm spin}$
are proportional to $\sin\theta$ and $\sin^2\theta$, respectively.
(We note that the conventional AHE (KL-term)
is given by replacing $\lambda l_+s_-$ and $h_{0-}s_+$
in Fig.  \ref{fig:sigmadiagram} (a) with $\lambda l_zs_z$ and $h_{0z}s_z$,
and $\sigma_{\rm AH}^{\rm KL} \sim h_{0z}\lambda/\Delta$.)

In $\rm Nd_2Mo_2O_7$, the relation $|\theta|\sim O(10^{-2})$ is realized.
Thus, we obtain $\sigma_{\rm AH}^{\rm spin}/\sigma_{\rm AH}^{\rm orb}\sim \theta$ 
since $\lambda \sim h_0 \sim 1000$K.
Therefore, $\sigma_{\rm AH}^{\rm orb}$ is about 100 times larger than 
$\sigma_{\rm AH}^{\rm spin}$ in $\rm Nd_2Mo_2O_7$. 
This result is recognized in the present numeraical calculation
in Fig. \ref{fig:theta2} and in Ref. \cite{TomizawaKontani}.

In $\rm Pr_2Ir_2O_7$, the relation $|\theta|\sim O(1)$ is realized.
Thus, we obtain $\sigma_{\rm AH}^{\rm spin}/\sigma_{\rm AH}^{\rm orb}\sim 
h_0/\lambda \sim O(10^2)$ since $h_{0}\sim 10$K and $\lambda\sim 3000$K. 
In the present numerical study, however, $\sigma_{\rm AH}$ is only 
$10\sim20$ times larger than $\sigma_{\rm AH}^{\rm spin}$, 
as shown in Fig. \ref{fig:thetaPrIr}.
This discrepancy originates from the higher-order correction 
of large $\lambda$ on the band-splitting $\Delta$:
In fact, $\sigma_{\rm AH}^{\rm orb}$ starts to decrease for $\lambda>600$K
since $\Delta$ increases with $\lambda$.

Finally, we comment that $\sigma_{\rm AH}^{\rm orb}$ is not
suppressed by large crystalline electric field.
Since $\hat{l}_\pm\hat{s}_\mp$ mixes the states
$|a_{1g},\sigma\rangle$ and $|e'_g,-\sigma\rangle$,
$\sigma_{\rm AH}^{\rm orb}$ will be large if these two states
occupy large portion of the DOS at the Fermi level. 
This situation is actually realized the presence of crystalline electric field,
as shown in Fig. \ref{fig:DOS} (b).
For this reason, large $\theta$-linear spin structure-driven AHE is realized
for $E_0=-2$.

\subsection{Summary}

In summary, we studied the AHE in the pyrochlore type $t_{2g}$-orbital model 
in the presence of non-collinear magnetic configurations and the crystalline electric field. 
Thanks to the SOI, the complex $d$-orbital wave function
is modified by the tilting angle $\theta$,
and the resultant orbital AB phase gives large $\theta$-linear AHC.
This orbital term, $\sigma_{\rm AH}^{\rm orb}$, dominates the AHE in 
$\rm Nd_2Mo_2O_7$ since the spin chirality term,
$\sigma_{\rm AH}^{\rm spin}$, is proportional to $\theta^2\ (\ll1)$.
The obtained numerical results are qualitatively equal to the results
in Ref. \cite{TomizawaKontani}.

In $\rm Pr_2Ir_2O_7$, $\sigma_{\rm AH}^{\rm orb}$ also dominates 
$\sigma_{\rm AH}^{\rm spin}$ since the SOI for Ir 5$d$-electron
($\lambda\sim3000$K) is much larger than the $d$-$f$ exchange interaction
($J_{df}\sim20$K).
In particular, the present orbital mechanism can explain the 
``hysteresis behavior of the AHC'' or 
``residual AHC under zero magnetization''
reported in $\rm Pr_2Ir_2O_7$ below $T^H\approx1$K,
if we assume small magnetic order of Ir $5d$-electrons at $T^H$.
In fact, the AHC under the Ir spin structure in Fig. \ref{fig:thetaPrIr} (c),
which would be realized below $T_{\rm N}^{\rm Ir}$ under 
weak exchange field from ``2in 2out'' Pr order,
is finite as shown in Fig. \ref{fig:thetaPrIr} (a).
The total AHC will be insensitive to the formation of domain structure 
with three ``2in 2out'' Pr orders in Fig. \ref{fig:Pr-structure},
since $\sigma_{\rm AH}$'s due to three $[2\downarrow2\uparrow]$ structures
are equivalent.
The AHC obtained in the present study
is expected to give a major part of the 
AHC observed in three dimensional compounds,
as discussed in Appendix C.

Since  $\sigma_{\rm AH}^{\rm orb}$ in the present model
is nonzero unless $\bm{n}_{A}||\bm{n}_{B}||\bm{n}_{C}$,
the realization condition for the orbital mechanism
is just the ``non-collinearity of the spin structure'',
which is much more general than that for $\sigma_{\rm AH}^{\rm spin}$.
The orbital mechanism might be the origin of interesting
spin structure-driven AHE in Fe$_3$Sn$_2$ \cite{Fenner,Kida}
and PdCrO$_2$ \cite{Takatsu}.

\begin{acknowledgments}
The authors are grateful to M. Sato, Y. Yasui, D. S. Hirashima,
Y. Maeno, H. Takatsu, S. Nakatsuji and Y. Machida for fruitful discussions.
This work has been supported by a Grant-in-Aid for Scientific Research
on Innovative Areas ``Heavy Electrons'' (No. 20102008) of
The Ministry of Education, Culture, Sports, Science, and Technology, Japan.

\end{acknowledgments}

\appendix

\section{Hopping integral between the sites with the different coordinates}
In this Appendix, we derive the hopping integrals between the sites 
with the different $d$-orbital coordinates as shown in Fig. \ref{fig:pyroxyz2}.
Here, we represent the five $d$-orbitals $xy$, $yz$, $zx$, $x^2-y^2$ and $3z^2-r^2$ as 1, 2, 3, 4 and 5. 
The wavefunctions in the $d$-orbitals are given by
\begin{eqnarray*}
\phi_{1}&=&\frac{1}{\sqrt{2}i}(Y_2^{2}-Y_2^{-2})=A\frac{xy}{r^2}\\
\phi_{2}&=&\frac{-1}{\sqrt{2}i}(Y_2^{1}+Y_2^{-1})=A\frac{yz}{r^2}\\
\phi_{3}&=&\frac{-1}{\sqrt{2}}(Y_2^{1}-Y_2^{-1})=A\frac{zx}{r^2}\\
\phi_{4}&=&\frac{1}{\sqrt{2}}(Y_2^{2}+Y_2^{-2})=\frac{1}{2}A\frac{x^2-y^2}{r^2}\\
\phi_{5}&=&Y_2^0=\frac{\sqrt{3}}{2}A\frac{3z^2-r^2}{r^2}
\end{eqnarray*}
where $Y_l^m$ is the spherical harmonics and $A=\sqrt{15/4\pi}$.

We consider the coordinate transformation matrix $\hat{O}^{\rm AB}$,
which transforms $(n_{x}, n_{y}, n_{z})_{\rm B}$ in the $(xyz)_{\rm B}$-coordinate
into $(n_{x}, n_{y}, n_{z})_{\rm A}$ in the $(xyz)_{\rm A}$-coordinate 
as $(n_{x}, n_{y}, n_{z})_{\rm A}\hat{O}^{\rm AB}=(n_{x}, n_{y}, n_{z})_{\rm B}$.
It is given by 
\begin{eqnarray}
\hat{O}^{\rm AB}=\frac{1}{9}
\begin{pmatrix}
-4&8&-1\\
-7&-4&4\\
-4&-1&8
\end{pmatrix}.
\end{eqnarray}
Since $\hat{O}^{\rm BC}$ and $\hat{O}^{\rm CA}$ are equivalent to $\hat{O}^{\rm AB}$, 
we have to derive only the hopping integral between sites A and B. 
Using $r^{\rm B}_{l'}=r^{\rm A}_lO^{\rm AB}_{ll'}$ where $l,l'=x,y,z$, 
the wavefunction for orbital $\beta$ at site B can be expressed 
as linear combination of the wavefunction for orbital $\gamma$ at site A. 
Thus, 
\begin{equation}
\phi_{{\rm B} \beta}=\sum_{\gamma} a_{\rm AB}(\beta,\gamma)\phi_{{\rm A} \gamma},
\end{equation}
where
\begin{eqnarray*}
a_{\rm AB}(1,1)&=&O^{\rm AB}_{11}O^{\rm AB}_{22}+O^{\rm AB}_{12}O^{\rm AB}_{21},\\
a_{\rm AB}(1,2)&=&O^{\rm AB}_{12}O^{\rm AB}_{23}+O^{\rm AB}_{13}O^{\rm AB}_{21},\\
a_{\rm AB}(1,3)&=&O^{\rm AB}_{11}O^{\rm AB}_{23}+O^{\rm AB}_{13}O^{\rm AB}_{22},\\
a_{\rm AB}(1,4)&=&2O^{\rm AB}_{11}O^{\rm AB}_{21}+O^{\rm AB}_{13}O^{\rm AB}_{23},\\
a_{\rm AB}(1,5)&=&\sqrt{3}O^{\rm AB}_{13}O^{\rm AB}_{23},
\end{eqnarray*}
\begin{eqnarray*}
a_{\rm AB}(2,1)&=&O^{\rm AB}_{21}O^{\rm AB}_{32}+O^{\rm AB}_{22}O^{\rm AB}_{31},\\
a_{\rm AB}(2,2)&=&O^{\rm AB}_{22}O^{\rm AB}_{33}+O^{\rm AB}_{23}O^{\rm AB}_{32},\\
a_{\rm AB}(2,3)&=&O^{\rm AB}_{21}O^{\rm AB}_{33}+O^{\rm AB}_{23}O^{\rm AB}_{31},\\
a_{\rm AB}(2,4)&=&2O^{\rm AB}_{21}O^{\rm AB}_{31}+O^{\rm AB}_{23}O^{\rm AB}_{33},\\
a_{\rm AB}(2,5)&=&\sqrt{3}O^{\rm AB}_{23}O^{\rm AB}_{33},
\end{eqnarray*}
\begin{eqnarray*}
a_{\rm AB}(3,1)&=&O^{\rm AB}_{11}O^{\rm AB}_{32}+O^{\rm AB}_{12}O^{\rm AB}_{31},\\
a_{\rm AB}(3,2)&=&O^{\rm AB}_{12}O^{\rm AB}_{33}+O^{\rm AB}_{13}O^{\rm AB}_{32},\\
a_{\rm AB}(3,3)&=&O^{\rm AB}_{11}O^{\rm AB}_{33}+O^{\rm AB}_{13}O^{\rm AB}_{31},\\
a_{\rm AB}(3,4)&=&2O^{\rm AB}_{11}O^{\rm AB}_{31}+O^{\rm AB}_{13}O^{\rm AB}_{33},\\
a_{\rm AB}(3,5)&=&\sqrt{3}O^{\rm AB}_{13}O^{\rm AB}_{33}.
\end{eqnarray*}
Therefore, the hopping integral $t_{{\rm B}\beta,{\rm A}\alpha}({\boldsymbol R}_{\rm AB})=\langle {\rm B}\beta, {\boldsymbol R}_{\rm B}|H_0|{\rm A}\alpha, {\boldsymbol R}_{\rm A}\rangle$ is given by
\begin{equation}
t_{{\rm B}\beta,{\rm A}\alpha}({\boldsymbol R}_{\rm AB})=\sum_{\gamma}a_{\rm AB}(\beta,\gamma)\tilde{t}_{{\rm A}\gamma,{\rm A}\alpha}({\boldsymbol R}_{\rm AB}),
\end{equation}
where $\tilde{t}_{{\rm A}\gamma,{\rm A}\alpha}({\boldsymbol R}_{\rm AB})=\langle {\rm A}\gamma, {\boldsymbol R}_{\rm B}|H_0|{\rm A}\alpha, {\boldsymbol R}_{\rm A}\rangle$ 
is the usual hopping integral between the equivalent coordinates, which is given by the SK parameter table in Ref. \cite{SlaterKoster}. 

\section{Local effective field from $\rm Pr$ tetrahedron}
In this Appendix, we derive the local effective field at 
$\rm Ir$ sites induced by the spin structure of $\rm Pr$ tetrahedron.
Sites A, B, C and D of Pr tetrahedron are located at 
$(1/4,0,0)$, $(0,1/4,0)$, $(0,0,1/4)$ and $(1/4,1/4,1/4)$, respectively, 
in the $xyz$-coordinate as shown in Table I, and 
the center of the tetrahedron is located at $(1/8,1/8,1/8)$. 

Under the strong field along $[111]$($>\!\!>0.7 {\rm Tesla}$), 
the spins of Pr tetrahedron form ``3in 1out" structure. 
The spin configurations at A$\sim$D are given by 
\begin{eqnarray*}
{\boldsymbol P}_{\rm A}^{\rm 3\downarrow1\uparrow}&=&(-1,1,1)/\sqrt{3},\\
{\boldsymbol P}_{\rm B}^{\rm 3\downarrow1\uparrow}&=&(1,-1,1)/\sqrt{3},\\
{\boldsymbol P}_{\rm C}^{\rm 3\downarrow1\uparrow}&=&(1,1,-1)/\sqrt{3},\\
{\boldsymbol P}_{\rm D}^{\rm 3\downarrow1\uparrow}&=&(1,1,1)/\sqrt{3}.
\end{eqnarray*}
In the intermediate field($\sim 0.7{\rm Tesla}$), the spins of Pr tetrahedron
is expected to form three kinds of ``2in 2out" structures which have 
negative Zeeman energy. 
First, we consider the case in which only one of three ``2in 2out" 
structures is realized. 
We choose one of three ``2in 2out" structures, which is obtained by 
inverting only ${\boldsymbol P}_{\rm C}^{\rm 3\downarrow1\uparrow}$ 
in the ``3in 1out'' structure.
That is, the configuration of the Pr spins in this 
``2in 2out" structure is given by 
${\boldsymbol P}_{\rm A}^{\rm 2\downarrow2\uparrow}={\boldsymbol P}_{\rm A}^{\rm 3\downarrow1\uparrow}$, 
${\boldsymbol P}_{\rm B}^{\rm 2\downarrow2\uparrow}={\boldsymbol P}_{\rm B}^{\rm 3\downarrow1\uparrow}$, 
${\boldsymbol P}_{\rm C}^{\rm 2\downarrow2\uparrow}=-{\boldsymbol P}_{\rm C}^{\rm 3\downarrow1\uparrow}$ and 
${\boldsymbol P}_{\rm D}^{\rm 2\downarrow2\uparrow}={\boldsymbol P}_{\rm D}^{\rm 3\downarrow1\uparrow}$.
We also consider another case where three ``2in 2out" structure are averaged. 
Then, the Pr moments at sites A, B and C are given by 
$1/3$ of the Pr moments in ``3in 1out" structure.
That is, the Pr spin in the averaged ``2in 2out" structures is given by 
${\boldsymbol P}_{\rm A}^{\overline{\rm 2\downarrow2\uparrow}}={\boldsymbol P}_{\rm A}^{\rm 3\downarrow1\uparrow}/3$, 
${\boldsymbol P}_{\rm B}^{\overline{\rm 2\downarrow2\uparrow}}={\boldsymbol P}_{\rm B}^{\rm 3\downarrow1\uparrow}/3$, 
${\boldsymbol P}_{\rm C}^{\overline{\rm 2\downarrow2\uparrow}}={\boldsymbol P}_{\rm C}^{\rm 3\downarrow1\uparrow}/3$ and 
${\boldsymbol P}_{\rm D}^{\overline{\rm 2\downarrow2\uparrow}}={\boldsymbol P}_{\rm D}^{\rm 3\downarrow1\uparrow}$.

The effective magnetic fields at Ir sites are obtained by summing six Pr spins:
Ir$_i$ atom is surrounded by two Pr$_{j}$, two Pr$_{k}$ and  two Pr$_{l}$, 
where we represent $\{i,j,k,l\}$ as a permutation of sites $\{\rm A, B, C, D\}$. 
Therefore, the local exchange fields at Ir sites are given by
\begin{eqnarray}
{\boldsymbol h}_i^{n\downarrow m\uparrow}=-J_{df}\sum_{j\neq i}2{\boldsymbol P}_j^{n\downarrow m\uparrow}.
\label{eq:h}
\end{eqnarray}
where $J_{df}$ is $d$-$f$ exchange interaction.
Hereafter, we assume $J_{df}>0$. 
We calculate the local exchange field ${\boldsymbol h}$ using above equation. 
The obtained results in each Pr structure are as follows: 
In the ``3in 1out" structure case,
\begin{eqnarray*}
{\boldsymbol h}_{\rm A}^{\rm 3\downarrow1\uparrow}&=&-\tilde{J}(3,1,1),\\
{\boldsymbol h}_{\rm B}^{\rm 3\downarrow1\uparrow}&=&-\tilde{J}(1,3,1),\\
{\boldsymbol h}_{\rm C}^{\rm 3\downarrow1\uparrow}&=&-\tilde{J}(1,1,3).\\
{\boldsymbol h}_{\rm D}^{\rm 3\downarrow1\uparrow}&=&-\tilde{J}(1,1,1).
\end{eqnarray*}
where $\tilde{J}=2J_{df}/\sqrt{3}$. 
In the ``2in 2out" structure case, 
\begin{eqnarray*}
{\boldsymbol h}_{\rm A}^{\rm 2\downarrow2\uparrow}&=&-\tilde{J}(1,-1,3),\\
{\boldsymbol h}_{\rm B}^{\rm 2\downarrow2\uparrow}&=&-\tilde{J}(-1,1,3),\\
{\boldsymbol h}_{\rm C}^{\rm 2\downarrow2\uparrow}&=&-\tilde{J}(1,1,3).\\
{\boldsymbol h}_{\rm D}^{\rm 2\downarrow2\uparrow}&=&-\tilde{J}(-1,-1,3).
\end{eqnarray*}
In the averaged ``2in 2out" structure case,
\begin{eqnarray*}
{\boldsymbol h}_{\rm A}^{\overline{\rm 2\downarrow2\uparrow}}&=&-\tilde{J}(5/3,1,1),\\
{\boldsymbol h}_{\rm B}^{\overline{\rm 2\downarrow2\uparrow}}&=&-\tilde{J}(1,5/3,1),\\
{\boldsymbol h}_{\rm C}^{\overline{\rm 2\downarrow2\uparrow}}&=&-\tilde{J}(1,1,5/3).\\
{\boldsymbol h}_{\rm D}^{\overline{\rm 2\downarrow2\uparrow}}&=&-\tilde{J}(1,1,1)/3.
\end{eqnarray*}
Next, we rewrite the obtained ${\bm h}$'s in the $XYZ$-coordinate 
using $[n_X,n_Y,n_Z]=(n_x,n_y,n_z)\hat{O}^{-1}$, where the transformation matrix 
$\hat{O}^{-1}$ is given by 
\begin{eqnarray}
\hat{O}^{-1}=\frac{1}{\sqrt{6}}
\begin{pmatrix}
-\sqrt{3}&-1&\sqrt{2}\\
\sqrt{3}&-1&\sqrt{2}\\
0&2&\sqrt{2}
\end{pmatrix}.
\end{eqnarray}
In the ``3in 1out" structure case, 
\begin{eqnarray*}
{\boldsymbol h}_{\rm A}^{\rm 3\downarrow1\uparrow}&=&-\tilde{J}[-2\sqrt{3},-2,5\sqrt{2}]/\sqrt{6},\\
{\boldsymbol h}_{\rm B}^{\rm 3\downarrow1\uparrow}&=&-\tilde{J}[2\sqrt{3},-2,5\sqrt{2}]/\sqrt{6},\\
{\boldsymbol h}_{\rm C}^{\rm 3\downarrow1\uparrow}&=&-\tilde{J}[0,4,5\sqrt{2}]/\sqrt{6},
 \\
{\boldsymbol h}_{\rm D}^{\rm 3\downarrow1\uparrow}&=&-\tilde{J}[0,0,3\sqrt{2}]/\sqrt{6}.
\end{eqnarray*}
In spherical coordinates, the direction of the local exchange fields at site A, B and C are 
$\theta=29.5^\circ$ and $(\phi_{\rm A},\phi_{\rm B},\phi_{\rm C})
=(-5\pi/6, -\pi/6, \pi/2)$.
In the ``2in 2out" structure case,
\begin{eqnarray*}
{\boldsymbol h}_{\rm A}^{\rm 2\downarrow2\uparrow}&=&-\tilde{J}[-2\sqrt{3},6,3\sqrt{2}]/\sqrt{6},\\
{\boldsymbol h}_{\rm B}^{\rm 2\downarrow2\uparrow}&=&-\tilde{J}[2\sqrt{3},6,3\sqrt{2}]/\sqrt{6},\\
{\boldsymbol h}_{\rm C}^{\rm 2\downarrow2\uparrow}&=&-\tilde{J}[0,4,5\sqrt{2}]/\sqrt{6},
\\
{\boldsymbol h}_{\rm D}^{\rm 2\downarrow2\uparrow}&=&-\tilde{J}[0,8,\sqrt{2}]/\sqrt{6},
\end{eqnarray*} 
that is,
$\theta_A=\theta_B=31.5^\circ$, $\theta_C=58.5^\circ$, and
$(\phi_{\rm A},\phi_{\rm B},\phi_{\rm C})
=(2\pi/3,\pi/3,\pi/2)$.
In the average ``2in 2out" structure case,
\begin{eqnarray*}
{\boldsymbol h}_{\rm A}^{\overline{\rm 2\downarrow2\uparrow}}&=&-\tilde{J}[-2\sqrt{3},-2,11\sqrt{2}]/3\sqrt{6},\\
{\boldsymbol h}_{\rm B}^{\overline{\rm 2\downarrow2\uparrow}}&=&-\tilde{J}[2\sqrt{3},-2,11\sqrt{2}]/3\sqrt{6},\\
{\boldsymbol h}_{\rm C}^{\overline{\rm 2\downarrow2\uparrow}}&=&-\tilde{J}[0,4,11\sqrt{2}]/3\sqrt{6}, \\
{\boldsymbol h}_{\rm D}^{\rm 3\downarrow1\uparrow}&=&-\tilde{J}[0,0,\sqrt{2}]/\sqrt{6},
\end{eqnarray*}
that is, 
$\theta=14.4^\circ$ and $(\phi_{\rm A},\phi_{\rm B},\phi_{\rm C})
=(-5\pi/6, -\pi/6, \pi/2)$.

\section{AHC in three dimensional compounds}
In this paper, we have studied the AHE in the kagome lattice model,
which represents the two-dimensional Mo or Ir network 
in the pyrochlore compounds.
In the presence of ``3in 1out'' or ``2in 2out'' of Nd or Pr 
spin-ice order, it was shown that prominent spin structure-driven AHE 
are induced on the kagome lattice on the [1,1,1] plane.
However, other three kagome layers on the [1,1,-1], [1,-1,1] 
and [-1,1,1] planes, which are not perpendicular to the magnetic field,
also give finite contribution to the AHC.

In this section, 
we shortly discuss the total AHC induced by four kagome lattices,
assuming that these lattices are independent.
Here, we put the magnetic field parallel to the $[1,1,1]$ plane,
which is given by the ABC plane in Fig. \ref{fig:pyroXYZ} (a),
and apply the electric field along Y axis.
Then, the AHC due to the [1,1,1] plane, $\sigma_{\rm AH}^{[1,1,1]}$,
is given in the present study.
Considering the relative angles and positions of other three kagome lattices, 
it is easy to show that the total AHC 
due to the electric field on the $[1,1,1]$ plane is given by
$\sigma_{\rm AH}^{\rm tot}=\sigma_{\rm AH}^{[1,1,1]}+
(\sigma_{\rm AH}^{[1,1,-1]}+\sigma_{\rm AH}^{[1,-1,1]}+\sigma_{\rm AH}^{[-1,1,1]})/3$.
Note that $[1,1,-1]$, $[1,-1,1]$, and $[-1,1,1]$ planes are respectively 
given by ABD, ACD, and BCD planes in Fig. \ref{fig:pyroXYZ} (a),
where D represents the apical Ir site.

First, we consider the AHC in Nd$_2$Mo$_2$O$_7$.
Under this $120^\circ$ structure of Ir spin (see Fig. \ref{fig:Pr-structure}),
the effective magnetic flux due to the orbital AB effect
is proportional to the tilting angle $\theta$; 
$\Phi^{[1,1,1]}=a\theta$.
However, the orbital AB phase for other kagome layers are different.
In the spinel-type kagome lattice studied in Ref. \cite{TomizawaKontani},
we can show that the effective magnetic flux for other layers are
$\Phi^{[1,1,-1]}=\Phi^{[1,-1,1]}=\Phi^{[-1,1,1]}=-(a/3)\theta$.
Then, the total AHC is given by
$\sigma_{\rm AH}^{\rm tot}=(2/3)\sigma_{\rm AH}^{[1,1,1]}$.

Next, we consider the AHC of Pr$_2$Ir$_2$O$_7$
under $[2\downarrow2\uparrow]_{H\rightarrow+0}$ Ir spin structure,
considering only the conventional KL term that is proportional to 
the perpendicular magnetization.
Then, it is easy to show that 
$\sigma_{\rm AH}^{[1,1,1]}=-\sigma_{\rm AH}^{[1,1,-1]}=\sigma_{\rm AH}^{[1,1,-1]}
=\sigma_{\rm AH}^{[-1,1,1]}$. 
Thus, the total AHC is given by
$\sigma_{\rm AH}^{\rm tot}=(4/3)\sigma_{\rm AH}^{[1,1,1]}$.

As a result, $\sigma_{\rm AH}^{[1,1,1]}$ gives the main contribution
to the total AHC in both cases.
Therefore, we expect that $\sigma_{\rm AH}^{[1,1,1]}$
studied in the present study represents a major part of the 
AHC observed in three dimensional compounds.



\end{document}